\documentclass[letterpaper, 10 pt, conference]{IEEEtran}  

\IEEEoverridecommandlockouts                              

\usepackage[utf8]{inputenc}
\usepackage{amsmath, amsfonts,amssymb,amsthm}
\usepackage{graphicx,psfrag,color,subfigure}
\usepackage{enumerate}
\newtheorem{problem}{Problem}

\newtheorem{theorem}{Theorem}

\newtheorem*{notation}{Notation}

\newtheorem{remark}{Remark}
\usepackage{algorithm,algorithmic}

\DeclareSymbolFont{extraup}{U}{zavm}{m}{n}
\DeclareMathSymbol{\vardiamond}{\mathalpha}{extraup}{87}

\usepackage{pgf,pgfplots,tikz}
\usetikzlibrary{arrows,automata,plotmarks}
\usetikzlibrary{positioning}
\tikzset{
    state/.style={
           rectangle,
           rounded corners,
           draw=black,
           minimum height=1em,
           inner sep=1pt,
           text centered,
           },
}

\title{\LARGE \bf
Optimal control for a robotic exploration, pick-up and delivery problem
}
\author{Vladislav Nenchev, Christos G. Cassandras, and J\"org Raisch%
\thanks{V.\,Nenchev is with the Control Systems Group, Technische Universit\"at Berlin, Germany. {Corresponding email: \tt\small nenchev@control.tu-berlin.de}}%
\thanks{C.\,G.\,Cassandras is with the Division of Systems Engineering and Center for Information and Systems Engineering, Boston University, MA, USA. {\tt\small cgc@bu.edu}}%
\thanks{J.\,Raisch is with the Control Systems Group, Technische Universit\"at Berlin and the Systems and Control Theory Group, Max Planck Institute for Dynamics of Complex Technical Systems, Magdeburg, Germany. {\tt\small raisch@control.tu-berlin.de}}%
\thanks{C.\,G.\,Cassandras was supported in part by NSF under grants CNS-1239021, ECCS-1509084, and IIP-1430145, by AFOSR under grant FA9550-15-1-0471, and by ONR under grant N00014-09-1-1051.}%
}

\begin{document}

\maketitle
\thispagestyle{empty}
\pagestyle{empty}

\begin{abstract}
This paper addresses an optimal control problem for a robot that has to find and collect a finite number of objects and move them to a depot in minimum time. The robot has fourth-order dynamics that change instantaneously at any pick-up or drop-off of an object. The objects are modeled by point masses with a-priori unknown locations in a bounded two-dimensional space that may contain unknown obstacles. For this hybrid system, an Optimal Control Problem (OCP) is approximately solved by a receding horizon scheme, where the derived lower bound for the cost-to-go is evaluated for the worst and for a probabilistic case, assuming a uniform distribution of the objects. First, a time-driven approximate solution based on time and position space discretization and mixed integer programming is presented. Due to the high computational cost of this solution, an alternative event-driven approximate approach based on a suitable motion parameterization and gradient-based optimization is proposed. The solutions are compared 
in a numerical example, suggesting that the latter approach offers a significant computational advantage while yielding similar qualitative results compared to the former. The methods are particularly relevant for various robotic applications like automated cleaning, search and rescue, harvesting or manufacturing.\\
\end{abstract}

\section{Introduction}

One of the major challenges in autonomous robotic navigation is coping with uncertainties arising from limited a-priori knowledge of the environment. Acquiring necessary information and achieving the overall goal are complementary subtasks that require adapting the motion of a robot during mission execution, typically accompanied by minimizing a performance criterion. In this work we address an Optimal Control Problem (OCP) for a robot with fourth-order dynamics that has to find, collect and move a finite number of objects to a designated spot in minimum time. The objects with a-priori known masses are located in a bounded two-dimensional space, where the robot is capable of localizing itself using a state-of-the-art simultaneous localization and mapping (SLAM) system \cite{Thrun2008}. The challenging aspects of the problem at hand are (at least) threefold. One of them arises due to the discontinuity of the value function denoting the overall completion time, which makes it hard to obtain an explicit 
controller even for deterministic linear systems \cite{Bryson1975,Clarke2010}. Fortunately, a wide range of approximate solutions has been proposed, including approaches based on numerical continuation \cite{Feng1986}, value set approximation \cite{Mitchell2005}, multi-parametric programming \cite{Grieder2005} etc. Another challenge follows from the requirement to collect a finite number of objects and drop them at a particular spot, both leading to autonomous switchings of the robot's continuous dynamics. While deterministic versions of this problem can be handled efficiently, e.g., by two-stage optimization \cite{Seatzu2006, Nenchev2015} or relaxation \cite{Passenberg2009}, the complexity of most approaches for stochastic setups scales poorly with the problem size \cite{Cassandras2010}. Since the robot has to reach the corresponding locations of the objects or the depot with minimal overall cost, the overall problem also contains an instance of the well-known NP-hard Traveling Salesperson Problem (TSP) \cite{Applegate2011}. Further, optimal exploration of a limited space is an inherently difficult problem by itself. Minimizing the expected time for detecting a target located on a real line with a known probability distribution by a searcher that can change its motion direction instantaneously, has a bounded maximal velocity and starts at the origin, was originally addressed in \cite{Bellman1963}. Different versions of this problem have received considerable attention from several research communities, e.g., as a ``pursuit-evasion game'' in game theory \cite{Guibas96, Alpern2003}, as a ``cow-path problem'' in computer science \cite{Kao1996} or as a ``coverage problem'' in control \cite{Cortes2004,Zhong2011}, but its solution for a general probability distribution or a general geometry of the region is, to a large extent, still an open question. Effective approaches for the related persistent monitoring problem based on estimation \cite{Leny2009}, linear programming \cite{Smith2012} or parametric optimization \cite{Cassandras2013} have been also been proposed. OCPs with uncertainties have also been addressed by certainty equivalent event-triggered \cite{Molin2013}, minimax \cite{Axelsson2008} and sampling-based \cite{Rickert2014} optimization schemes. While methods for Partially Observable Markov Decision Processes (POMDP's) can also be applied, e.g., \cite{Toussaint2012,Amato2015}, they typically become computationally infeasible for larger problem instances. Due to the aforementioned aspects, the problem at hand has exponential complexity in the number of objects and for any chosen time and space discretization. In this context, employing a discrete abstraction of the underlying continuous dynamics is often only possible by introducing a hierarchical decomposition \cite{Lahijanian2016}, or additional assumptions that simplify the implementation of automatically synthesized hybrid controllers \cite{Raman2015}. Alternatively, one may resort to receding horizon approaches that have been shown to outperform other optimization methods under the presence of uncertainty, e.g., for the elevator dispatching problem \cite{Wesselowski2006}, multi-agent reward collection problems \cite{Khazaeni2014} or planning with temporal logic constraints \cite{Nenchev2016}. 

For a scenario where the number of objects is finite but unknown, a combined optimal exploration and control scheme for a robot that has to find, collect and move objects in a two-dimensional position space was proposed in \cite{Nenchev2013}. The approach was based on a policy enforcing a pick-up upon an object's detection, followed by a certainty equivalent discrete optimization on a finite abstraction of the robot's motion in the environment. This heuristic restriction was omitted in \cite{Nenchev2014}, where optimal exploration and control solutions for the worst and a probabilistic case assuming a uniform distribution of the objects on a line interval were derived. Since a direct generalization of this result for higher dimensions was not possible, this paper proposes and compares two approximate receding horizon approaches. The first is based on discretizing time and space and solving a non-convex OCP over a finite horizon by a Mixed Integer Programming (MIP) implementation. In the second approach, the motion of the robot is parameterized by a finite number of parameters. This enables the use of Infinitesimal Perturbation Analysis (IPA) \cite{Cassandras2010a} to solve the worst and probabilistic case OCPs by a bi-level iterative optimization scheme, solved only whenever new information becomes available. Preliminary versions of these approaches along a fixed exploratory trajectory have been presented in \cite{Nenchev2015a}. Here we extend the methods such that the shape of the exploratory trajectory can be adjusted online, which is particularly useful under the presence of a-priori unknown obstacles. 

The remainder of the paper is organized as follows: in Sec.\,\ref{sec:prob}, we present the problem formulation. Sec.\,\ref{sec:solution} starts with a brief discussion on the performance index and introduces a lower bound for the cost-to-go, followed by the proposed time-driven (Sec.\,\ref{sec:space}) and event-driven approaches (Sec.\,\ref{sec:parameterize}). The four methods are then compared in a numerical example (Sec.\,~\ref{sec:example}), followed by the conclusions in Sec.\,\ref{sec:conclusion}.

\begin{notation}
For a set $S$, $|S|$ and $2^S$ denote its cardinality and the set of all of its subsets (power set), respectively. For $r\in \mathbb{R}$, respectively, $r\in \mathbb{R}^n$, $|r|$ and $\|r\|$ denote the absolute value and the Euclidean norm. $\mathbf{I}_n$ is an identity matrix with dimension $n$. $\mathbf{0}_{m,n}$ represents an $m\times n$ matrix with zero entries. For a vector of zeros or ones with length $m$, we write $\mathbf{0}_m$ or $\mathbf{1}_m$, respectively. $\mathbb{R},\mathbb{R}_{\geq0},\mathbb{R}_{>0}$ denote the sets of reals, non-negative reals and positive reals, respectively. We use the derivatives $\dot{x}(t)=\frac{d x(t)}{dt}$, $c'(s,\theta)=\frac{\partial c(s,\theta)}{\partial s}$ and the gradient $\nabla_\theta c(s,\theta)=\left[\frac{\partial c(s,\theta)}{\partial \theta_1},\ldots,\frac{\partial c(s,\theta)}{\partial \theta_n}  \right]^T$. 
\end{notation}

\section{Problem formulation}\label{sec:prob}

Consider a finite set of objects $O=\{o_1,\ldots,o_L\}$, where every $o_l,l\in\{1,\ldots,L\}$, is uniquely characterized by its position $p^{(l)}\in\mathcal{Y}_g$, $\mathcal{Y}_g=[-y_{\text{max}},y_{\text{max}}]\times [-y_{\text{max}},y_{\text{max}}]\subset\mathbb{R}^2$, and mass $m^{(l)}\in\mathbb{R}_{\geq 0}$. A robot has to find, collect and move all objects back to a designated spot (depot), located at $y_d=\mathbf{0}_2$, in minimum time. The robot is equipped with an omni-directional sensor footprint of size $r\ll y_{\text{max}}$ around its current position $y(t)\in\mathbb{R}^2$, hence covering the area
\begin{align}\label{sightreg}
\mathcal{O}(y(t))=\{y_p\in\mathbb{R}^2: \|y(t)-y_p\|\leq r\}.
\end{align}

The overall system is modeled by a hybrid automaton \cite{Henzinger2000}, i.e., a 9-tuple $\mathcal{H}=\{Q,\mathcal{X},F,U,E,\text{Inv},G,R,\text{Init}\}$. The discrete state at time $t$ is $q(t)=(q_1(t),q_2(t),q_3(t))$, where $q_1(t)\subseteq O$ is the set of objects being carried by the robot, $q_2(t)\subseteq O$ the set of objects that has been dropped at the depot prior to or at time $t$, and $q_3(t)\subseteq O$ is the set of objects that have been detected so far. Clearly, $q(t)\in Q$ with $Q\subseteq2^O\times 2^O\times 2^O$. The current mass of the robot is $m_q(t)=m_{\emptyset}+\sum_{l,o_l\in q_1} m^{(l)}$, where $m_{\emptyset}$ is the nominal mass of the robot. The overall continuous state $(x(t), \mathcal{Y}(t))\in \mathcal{X}$ consists of the robot state $x(t)=[y^T(t)\text{ }v^T(t)]^T\in X$, where $v(t)\in\mathbb{R}^2$ is the current velocity of the robot, and the region $\mathcal{Y}(t)\subseteq \mathcal{Y}_g$ that has not been explored at time $t$. The robot state $x(t)$ evolves according to a 
finite collection of vector fields $F=\{f_{q}\}_{q\in Q}$, i.e. 
\begin{align}\label{eq:system}
 \dot{x}(t)=& f_{q}(x,u)= \begin{bmatrix}
\mathbf{0}_{2,2} & \mathbf{I}_{2} \\
\mathbf{0}_{2,2} & \mathbf{0}_{2,2}\end{bmatrix}x(t)+ \frac{1}{{m_q(t)}}\begin{bmatrix}
\mathbf{0}_{2,2} \\
\mathbf{I}_{2}\end{bmatrix}u(t),
\end{align}
driven by the piecewise continuous control signal $u:[0,t_f]\to U:=\{\phi\in\mathbb{R}^2:\|\phi\|\leq 1\}$, where $t_f$ is the free final time for the overall assignment. As $Q$ is finite, the set of discrete state transitions (or events) $E\subseteq Q\times Q$ is also finite. Let $E$ be partitioned into $\Delta\cup\Pi\cup \Psi$, where for $q=(q_1,q_2,q_3),q'=(q_1',q_2',q_3')\in Q$,
\begin{align*}
\Delta=\{(q,q'): q_1'=q_1, q_2'=q_2, q_3'=q_3\cup\{o_l\},o_l\not\in q_3\}
\end{align*}
is the set of detection events,
\begin{align*}
 \Pi=\{(q,q'): q_1'\setminus q_1=\{o_l\},o_l\in q_3,q_2'=q_2,q_3'=q_3\},
\end{align*}
is the set of pick-up events, and
\begin{align*}
\Psi=\{(q,q'): q_1\neq \emptyset, q_1'=\emptyset, q_2'=q_2\cup q_1, q_3'=q_3\}
\end{align*}
corresponds to the set of drop-off events. With the introduced sensor paradigm \eqref{sightreg}, detection events occur when the distance between the current robot position and the position of an object that has not been detected so far becomes $r$. Pick-up events occur when the robot reaches the position of an object that has not been collected so far. Drop-off events occur when the robot reaches the depot and carries objects. In addition, for both pick-up and drop-off events, zero velocity is required. The corresponding conditions on $q$ and $x$ for the occurrence of detection, pick-up and drop-off events are captured by the invariant $\text{Inv}: Q \to 2^{X}$, i.e.,
\begin{align*}
\text{Inv}(q){=}\begin{cases}
X{\setminus}\{[y^T\text{ }v^T]^T: \|y-p^{(l)}\|{\neq}r\}, &\text{if } o_l\not\in q_3,\\
X{\setminus}\{[{p^{(l)}}^T\text{ }\mathbf{0}_{2}^T]^T\}, &\text{if } o_l{\not\in} q_1{\cup}q_2,\\
X{\setminus}\{\mathbf{0}_{4}\}, &\text{if } q_1\neq\emptyset,
              \end{cases}
              \end{align*}
               and the guard map $G: E\to 2^{X}$, i.e. with $e=(q,q')$,
              \begin{align*}
     G(e){=}\begin{cases}
\{[y^T\text{ }v^T]^T:\|y-p^{(l)}\|{=}r\},&\text{if } e{\in} \Delta, q_3'{\setminus} q_3{=}\{o_l\},\\
      \{[{p^{(l)}}^T\text{ }\mathbf{0}_{2}^T]^T\},&\text{if } e{\in} \Pi, q_1'{\setminus} q_1{=}\{o_l\},\\
\{\mathbf{0}_{4}\},&\text{if } e{\in} \Psi, q_1\neq \emptyset.
     \end{cases}         
\end{align*}
For example, upon a detection of a new object as per \eqref{sightreg}, when the robot is in the discrete state $q$, the first case of the Inv requires that a transition must occur, and the first case of $G$ allows a transition to a discrete state $q'$, where the discovered object is included in the detected objects set $q_3'$. The reset map $R:E\times X\to 2^X$ is trivial since no jumps of the continuous variables occur upon a discrete state switching. Note that the above conditions do not depend on $\mathcal{Y}(t)$, and hence, $\text{Inv}$, $G$ and $R$ map into $2^X$ instead of $2^\mathcal{X}$. As both the robot and the objects are represented by points in $\mathcal{Y}_g$, we assume that no collisions can occur. A practical setup that satisfies this assumption is, e.g., a quadrotor that has to explore a two-dimensional space on the ground from above. Finally, as the robot is assumed to start at the depot with zero velocity, and no objects have been detected, picked up or dropped off before that, the initial 
state set is $\text{Init}=\{(q(0),(x(0),\mathcal{Y}(0)))\}=\{((\emptyset,\emptyset,\emptyset),(\mathbf{0}_{4},\mathcal{Y}_g\setminus \mathcal{O}(\mathbf{0}_2))\}$. 
\begin{remark}
Obstacles in $\mathcal{Y}_g$ can be easily included in the proposed approaches. However, to keep notation as simple as possible, we omit their presence in the main analysis and briefly outline the solution that was used to handle the obstacle in the numerical example (Sec.\,VI) in a follow-up remark. 
\end{remark}

Solving the addressed problem involves $L$ detection events, $L$ pick-up events and up to $L$ drop-off events, as it can be advantageous to collect several objects on the way and drop them off simultaneously at the depot. Hence, for the total number $N$ of events, $2L<N\leq 3L$ holds. The time of the occurrence of event $n$, $1\leq n\leq N$ is denoted by $t_n$, $t_0$ is the initial time, $t_f=t_N$ the final time, and $t_0\leq t_1\leq\ldots\leq t_N$. The $N$ time intervals $\tau_n:=[t_{n{-}1},t_n]$, $n=1,\ldots,N$ form the time axis from the initial to the final time with $\tau:=(\tau_1,\ldots,\tau_N)$. The input is an ordered set of functions $u=(u^1,\ldots,u^{N})$, where $u^n:\tau_n\to U$ are absolutely continuous functions for $n\in\{1,\ldots,N\}$. Thus, if $\zeta=(\tau,q,\xi)_u$ is an execution of the hybrid automaton $\mathcal{H}$ for an input signal $u$, i.e. $(\tau,q,\xi)_u\vDash\mathcal{H}$, $q=(q^1,\ldots,q^N)$ is a discrete state trajectory with $q^n:\tau_n\to Q, q^n=\text{const}$. $\xi=(\xi^1,\
ldots,\
xi^N)$ is the continuous state trajectory with $\xi^n=(x^n,\mathcal{Y}^n)$, where $x^n:\tau_n\to X$ are absolutely continuous functions, and $\mathcal{Y}^n:\tau_n\to 2^{\mathcal{Y}_g}$ non-increasing functions, i.e., $\mathcal{Y}^n(t')\subseteq \mathcal{Y}^n(t)$ for $t\leq t'$. The cost of an execution is the total task time
\begin{align}\label{cost}
  t_f=\sum_{n=1}^{N} (t_{n}-t_{n{-}1})=t_N-t_0.
\end{align}

Let $\text{Fin}=\{(q_f=(\emptyset,O,O),(\mathbf{0}_4,\mathcal{Y}_g\setminus \cup_{\tilde{t}\in[0,t_f]}\mathcal{O}(y(\tilde{t}))))\}$ denote the set of states that can be reached upon completing the task. One way to account for the uncertainty in the addressed OCP is to minimize, at time $t$, the largest cost that may occur for a possible configuration of all objects that have not been discovered so far. Alternatively, the positions of the objects that have not been detected so far can be assumed to be independent identically distributed random variables with probability density functions 
\begin{align}\label{eq:pdf}
\mathcal{P}(p^{(l)}){=}\begin{cases}
         \frac{1}{\kappa(t)}, &\text{ if } p^{(l)}\in\mathcal{Y}(t),\\
         0, &\text{ if } p^{(l)}\in\mathcal{Y}_g\setminus\mathcal{Y}(t),
        \end{cases}
\end{align}
$\forall l$, where $\kappa(t)$ measures the size of $\mathcal{Y}(t)$. This leads to the following worst-case (A) and probabilistic (B) OCPs.

\begin{problem}\label{prob} 
At state $(q(t),\xi(t))$, find the input signal $u|_{[t,t_f]}$ for $\mathcal{H}$, such that for $p=\{p^{(l)}:o_l\not\in q_3(t)\}$
\begin{align*}
&A) \min_{u|_{[t,t_f]}} \max_{p} \text{ } (t_f-t),\text{ s.t. } (q(t_f),(x(t_f),\mathcal{Y}(t_f)))\in\text{Fin};\\
&B) \min_{u|_{[t,t_f]}} \text{ }E\{ t_f-t\},\text{ s.t. } (q(t_f),(x(t_f),\mathcal{Y}(t_f)))\in\text{Fin}.
\end{align*}
\end{problem}
Note that Problem~A is always deterministic, while Problem~B is probabilistic until the last detection of an object.

The outline of the solution reads as follows. First, we provide a discussion on the time-optimal value function and derive a lower bound for the cost-to-go. Then, we propose two approximation-based approaches for Problems A and B -- one that requires time discretization and re-computation at every time step, and one based on motion parameterization that allows for an event-driven implementation, i.e., the corresponding OCPs are re-solved only upon the occurrence of a detection event.

\section{Preliminary analysis}\label{sec:solution}

Let $t'\in(0,t_f]$ be a time instant at which the robot has reached a pick-up or drop-off location with zero velocity. The overall cost-to-go at state $(q(t),\xi(t)), t\in[0,t')$ is
\begin{align}\label{eq:detcost}
\begin{aligned}
 J((q(t),\xi(t)),u|_{[t,t_f]},p)=\ell(q(t),\xi(t),u|_{[t,t']},p){+}\\
 J(q(t'),\xi(t'),u|_{[t',t_f]},p),
 \end{aligned}
\end{align}
i.e., the sum of the time $\ell$ until the next pick-up or drop-off at time $t'$, and the remaining time until the final state is reached.
\begin{figure}
\begin{center}
\includegraphics{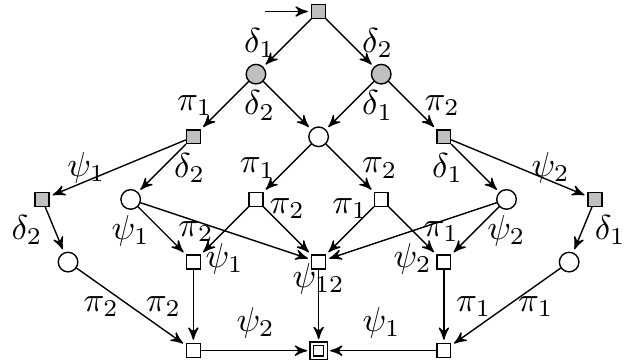}
\end{center}
\caption{Discrete dynamics of $\mathcal{H}$ for $O{=}\{o_1,o_2\}$ with $\delta_l\in\Delta$, $\pi_l\in\Pi$, $\psi_l\in\Psi$, $l\in\{1,2\},\psi_{12}\in\Psi$. Exploration takes place at gray states. The robot has zero velocity at states denoted by a square.}
\label{fig:graph}
\end{figure}
Assuming that all objects have been detected prior to $t'$, the second term on the right hand sight of \eqref{eq:detcost} is the cost of the optimal sequence of pick-ups and drop-offs, necessary for completing the overall task. Let the set of all corresponding discrete state strings from the state $q(t')=q$ to the final discrete state $q_f$ be denoted by 
\begin{align}\label{stringset}
\begin{aligned}
\Sigma_q:=\{&\sigma=q_0q_1\ldots q_d: (q_{i-1},q_{i}) \in(\Pi\cup\Psi),\\
&i\in\{1,\ldots,d\}, q_d=q_f, q_0=q\}. 
\end{aligned}
\end{align}
Minimizing the cost $J_\sigma$ of a particular sequence $\sigma\in\Sigma_q$ can be decoupled in terms of the input $u|_{[t',t_f]}$ at every pick-up and drop-off time instant $t_{i-1},t_i\in[t',t_f]$, i.e.,
\begin{align}\label{eq:sumcost}
\begin{aligned}
J^*_\sigma(q(t'),\xi(t'),p)&{=}\min_{u|_{[t',t_f]}}J_\sigma(q(t'),\xi(t'),u|_{[t',t_f]},p)\\
&{=}\sum_{i=1}^{d}\min_{u|_{[t_{i-1},t_{i}]}} J(q_{i-1},\xi(t_{i-1}),u|_{[t_{i-1},t_{i}]},p)
\end{aligned}
\end{align}
with $t_0=t'$ and $t_d=t_f$. Assuming the absence of obstacles in $\mathcal{Y}_g$, the time-optimal motion of the robot with dynamics \eqref{eq:system} from the hybrid state $(q_{i-1},(x(t_{i-1})=[y(t_{i-1})^T\text{ }\mathbf{0}_{2}^T]^T,\mathcal{Y}(t_{i-1})))$ to $(q_{i},(x(t_{i})=[y(t_{i})^T\text{ }\mathbf{0}_{2}^T]^T,\mathcal{Y}(t_{i})))$ with $y(t_i),y(t_{i-1})\in p\cup\{\mathbf{0}_2\}$ is on straight lines. Thus, using an affine transformation, \eqref{eq:system} can be reduced to a double integrator in one dimensional space. The OCP for the reduced model corresponds to the classical linear time-OCP \cite{Bryson1975} solved by a piecewise constant control that takes values in the set $\{\pm 1\}$ and yields the optimal cost $t_{i}-t_{i-1}=2\sqrt{m_{q_{i-1}}\|y(t_{i}){-}y(t_{i-1})\|}$. The controller can be transformed back to \eqref{eq:system} by using the inverse affine transformation (details can be found in \cite{Nenchev2015}). Since a transition from a hybrid state, where the robot with dynamics \eqref{
eq:system} has zero velocity, to another hybrid state, where the robot has zero velocity, can be tightly lower bounded by the cost for the time-optimal point-to-point motion of a double integrator with zero initial and final velocity, for the optimal cost of a string we obtain
\begin{align}\label{esttime}
\begin{aligned}
J^*_\sigma(q(t'),\xi(t'),p)=\sum_{i=1}^{d} 2\sqrt{m_{q_{i-1}}\|y(t_{i}){-}y(t_{i-1})\|}.
\end{aligned}
\end{align}
To illustrate this expression, consider a scenario with two remaining objects, both to be picked up and dropped. The corresponding discrete dynamics of $\mathcal{H}$ are captured by the quadruple $(Q,E,q_0,q_f)$ (Fig.\,\ref{fig:graph}), where $Q$ and $E$ are the corresponding sets of $\mathcal{H}$, and $q_0$ and $q_f$ are the initial and final discrete state (specified by Init and Fin), respectively. If both objects have been detected and the robot is at rest, the right hand side of \eqref{costest} denotes the actual cost-to-go for completing the task. In addition, the right hand side of \eqref{costest} can be used as a lower bound for the cost-to-go at time $t'$, where the robot is at rest but not both objects have been discovered, i.e.,
\begin{align}\label{costest}
\begin{aligned}
&J((q(t'),\xi(t')),u|_{[t',t_f]},p)\geq\\
&J_{lb}(q(t'),\xi(t'),p)=\min_{\sigma\in\Sigma_q} J^*_\sigma(q(t'),\xi(t'),p),
\end{aligned}
\end{align}
which represents the cost-to-go without taking into account exploration. 

\begin{figure}
\centering
\subfigure[]{
\includegraphics{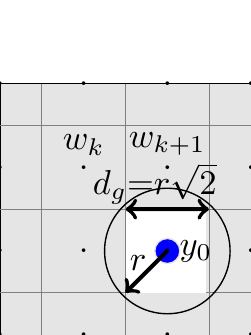}
 \label{fig:disc10}
}
\subfigure[]{
\includegraphics{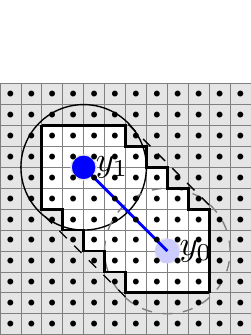}
 \label{fig:disc1}
}
\subfigure[]{
\includegraphics{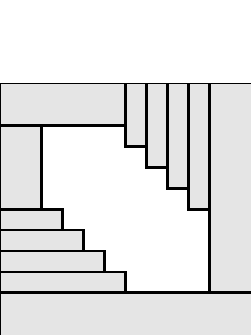}
 \label{fig:disc2}
}
\caption{A robot with sensing radius $r$ over the coarsest allowed grid (a). A snapshot of the robot that has moved from $y_0=\mathbf{0}_2$ to $y_1\neq \mathbf{0}_2$ with $d_g< r\sqrt{2}$ (b). The area covered along the path is under-approximated over the grid . The over-approximation $\tilde{\mathcal{Y}}(t)$ (in gray) of $\mathcal{Y}(t)$ is described by a finite number of rectangular regions (c) used for the time-driven optimization.}
\label{fig:discr}
\end{figure}

To obtain a finite conservative approximation for $\mathcal{Y}(t)$, introduce a finite cover of $\mathcal{Y}_g$ by cells $\omega_k,k\in\{1\ldots,K\}$ defined by a set of grid points $W=\{w_1,\ldots,w_K\}$, equally spaced by $d_g\leq r\sqrt{2}$, such that $\omega_k=\{y\in\mathcal{Y}_g:\|y-w_k\|_{\infty}\leq d_g/2\}$ (see Fig.\,\ref{fig:discr} for an example). Let $\mathcal{W}(t)$ denote the set of grid points, whose associated cells have not been completely covered by the robot's sensing range \eqref{sightreg} until time $t$, i.e. $\mathcal{W}(t)=\{w_i\in W: \omega_i\not\subseteq \cup_{\tilde{t}\in[0,t]} \mathcal{O}(y(\tilde{t}))\}$.  Thus, $\mathcal{Y}(t)$ is over-approximated by $\tilde{\mathcal{Y}}(t)=\{\cup_i \omega_i:w_i\in\mathcal{W}(t)\}$. With that, we can turn to approximate solutions of Problems A and B.

\section{Time-driven optimization}\label{sec:space}

In this section, we present an approximation of Problem~\ref{prob} based on equidistant time discretization.

\subsection{Worst-case solution}

By applying the min-max inequality and \eqref{costest}, since $\ell \geq 0$, $J_{lb}\geq 0$, for the (certainty equivalent) worst-case evaluation of \eqref{eq:detcost}, we obtain
\begin{align}\label{eq:costwc}
\begin{aligned}
 t_f^{\text{w}}-t&{=}\min_{u|_{[t,t_f]}} \max_{p} \text{ } J((q(t),\xi(t)),u|_{[t,t_f]},p)\\
 &{\geq} \max_{p} ((\min_{u|_{[t,t']}} \ell(q(t),\xi(t),u|_{[t,t']},p)){+}J_{lb}(q(t'),\xi(t'),p))\\ 
 &{\geq} \min_{u|_{[t,t']}} \text{ } \ell(q(t),\xi(t),u|_{[t,t']},p^*){+}J_{lb}(q(t')\xi(t'),p^*),
 \end{aligned}
\end{align}
where $p^*$ is defined as 
\begin{align}\label{nonconv}
\begin{aligned}
p^*=\arg \max_{p\in\tilde{\mathcal{Y}}(t)} J_{lb}(q(t')\xi(t'),p),
\end{aligned}
\end{align}
which follows from relaxing the assumption at time $t'$ in a sense that the robot has zero velocity at pick-up or drop-off locations, but not all objects have necessarily been detected before $t'$. Since $|\Sigma_q|$ is finite, it is possible to reformulate \eqref{nonconv} by introducing a dummy variable $\bar{t}\geq 0$ and $|\Sigma_q|$ additional nonlinear constraints for each string in $\Sigma_q$ leading to
\begin{align*}
\max_{\tilde{t},p\in\tilde{\mathcal{Y}}(t)} \bar{t}, \text{ s.t. } &\forall \sigma\in\Sigma_q, J^*_\sigma(q(t'),\xi(t'),p)-\bar{t}\geq 0.
\end{align*}

As the robot has zero velocity at $t_0=0$, the initial (approximately) optimal control $u|_{[0,t^{\text{w}}_f]}$ can be obtained by solving \eqref{nonconv} followed by re-translation to \eqref{eq:system}, as described in the previous section. Once the robot starts moving, optimizing the first term of the third line of \eqref{eq:costwc} at the optimum $p^*$ is difficult in continuous time. Therefore, consider a finite equidistant sampling of a time horizon beginning at $t$ with sampling time $t_s$, which we assume to include the yet unknown time $t'$, i.e. $i\in[0,N_{\text{max}}-1]$, $N_{\text{max}}\in\mathbb{N}$, $t=0,t'<(N_{\text{max}}-1)t_s$. Then, at every time instant, given the solution of \eqref{nonconv}, we solve
\begin{align}\label{eq:wcOCP}
\begin{aligned}
&\min_{u|_{[0,i-1]}} i t_s,\\
&\text{ s.t. } \eqref{nonconv}, \forall i\in[0,N_{\text{max}}-1], (q_i,\xi_i) \vDash \mathcal{C}(i),
\end{aligned}
\end{align}
where $\mathcal{C}(i)$ is the constraint set resulting from the corresponding discrete-time version of the hybrid automaton $\mathcal{H}$. The OCP can be approximately implemented as a MILP. For further implementation details, we refer the reader to Appendix~A and \cite{Nenchev2013} for a closely related OCP. 

\subsection{Probabilistic solution}

With \eqref{eq:pdf}, \eqref{eq:detcost} and \eqref{costest}, prior to the discovery of all objects, the optimal cost in the probabilistic case is given by the minimum expected time (omitting function arguments)
\begin{align}\label{stochint}
\begin{aligned}
E\{t_f{-}t\}{\geq}&\min_{u|_{[t,t_f]}} (E\{\ell\}+E\{J_{lb}\})\\
{\geq}&\underbrace{\frac{1}{\kappa}\min_{u|_{[t,t']}}\int_{\tilde{\mathcal{Y}}(t)} \ell d p}_{}+\underbrace{\frac{1}{\kappa}\int_{\tilde{\mathcal{Y}}(t)} \min_{\sigma\in\Sigma_q}J^*_\sigma dp}_{},\\[-1em]
&\geq\min_{u|_{[t,t']},p\in \tilde{\mathcal{Y}}(t)} \ell \quad \geq\frac{1}{\kappa}\min_{\sigma\in\Sigma_q}\int_{\tilde{\mathcal{Y}}(t)} J^*_\sigma dp
\end{aligned}
\end{align}
where $\kappa$ denotes the area of $\tilde{\mathcal{Y}}(t)$. The approximation of the first term follows from the fact that $E\{\ell\}$ is certainly greater or equal to the shortest time needed for the robot to move from its current position to a currently unexplored point in $\tilde{\mathcal{Y}}(t)$, while the approximation of the second term is obtained by applying Jensen's inequality. To compute the control $u|^*_{[t,t']}$ that minimizes the first term in \eqref{stochint}, we formulate a MILP analogously to \eqref{eq:wcOCP}. The second term in \eqref{stochint} is obtained through numerical integration of $J_\sigma$ for $p$ over $\tilde{\mathcal{Y}}(t)$, followed by choosing the sequence $\sigma^*\in\Sigma_q$ that yields the minimal cost. This allows for a receding horizon scheme that minimizes the cost-to-go at each time instant until all objects are dropped off. 

\section{Event-driven optimization}\label{sec:parameterize}

The approaches presented in the previous section require solving computationally expensive MIPs at each time instant online. Since the locations of the objects are the only source of uncertainty in the considered problem, the ultimate goal is a tractable and scalable, albeit suboptimal alternative that avoids time discretization and requires re-computation only upon a detection. The approach proposed in the following is based on restricting the motion of the robot to a pre-specified family of curves, whose shape is determined by a finite parameter vector, such that the cost-to-go can be evaluated efficiently. This allows for an event-driven scheme based on an iterative gradient-based optimization over the parameters of the curve only upon detection.

Let the robot's position be described by the parametric equation
\begin{equation}\label{parameterization}
 y(t)=c(s(t),\theta)=\begin{bmatrix}
                      c_1(s(t),\theta)\\
                      c_2(s(t),\theta)
                     \end{bmatrix}\in\mathbb{R}^2,
\end{equation}
where $s(t)$ denotes the position of the robot along the curve $c$, $\theta\in\mathbb{R}^a$ is a parameter vector that controls the shape of $c$, and $c$ is twice continuously differentiable with respect to $s$ and $\theta$. Let $\tilde{s}(t)$ be the monotonically non-decreasing curve length function of $c$ over $t\in[0,t_f]$. With $\alpha\in\mathbb{R}_{>0}$ denoting the arc-length of $c$, let $s(t)=\frac{\tilde{s}(t)}{\alpha}$ denote the normed arc-length variable, such that $s(0){=}0$ at the initial position, and $s(t_f)=1$ at the final position. The parametric functions employed in this work are Fourier series (see Appendix\,B) that exhibit rich expressiveness in terms of motion behaviors and allow for an efficient solution of the optimization problem. Other types of parametric functions or more complex robot dynamics may also be used, as outlined in Remark~\ref{otherparam}. 

Upon detection, optimization will be performed by a bi-level optimization algorithm, based on iteratively solving the following two OCPs:
\begin{enumerate}
 \item Find the parameter $\theta^*$ that determines the optimal shape of $c(s(t),\theta)$ solving Problem~1;
 \item Control the motion of the robot along $c(s(t),\theta^*)$ by the optimal $s^*(t)$ that respects the restrictions imposed by $\mathcal{H}$.
\end{enumerate}
The outline of the algorithm reads as follows: starting with an initial parameter guess for $\theta$, we solve the low-level OCP 2). Then, $\theta$ is updated by solving 1) using the solution of 2). The high-level OCP is solved by an augmented Lagrangian method that allows for replacing the constrained optimization problem by a series of unconstrained optimization problems. Employing Infinitesimal Perturbation Analysis \cite{Cassandras2010a}, we obtain the derivative of the augmented cost and solve the unconstrained OCPs by gradient-based methods. The steps 1) and 2) are solved iteratively until reaching a (local) minimum of the OCP, which is attained upon satisfying an iteration threshold condition. We start with solving the second step. 

\subsection{Optimal motion along the curve}

Let the first and second derivatives of \eqref{parameterization} w.r.t. $s$ be $c'(s,\theta)=\partial c/\partial s$ and $c''(s,\theta)=\partial^2 c/\partial s^2$, respectively. Further, let $\dot{s}=ds/dt$ and $\ddot{s}=d^2s/dt^2$ denote the time derivatives. For the velocity and the acceleration along \eqref{parameterization}, we respectively obtain
\begin{align*}
\begin{aligned}
 \dot{c}(s,\theta)=&\frac{d c(s(t),\theta)}{d t}=c'(s,\theta)\dot{s},\\
 \ddot{c}(s,\theta)=&\frac{d^2 c(s(t),\theta)}{d t^2}=c''(s,\theta)\dot{s}^2+c'(s,\theta)\ddot{s},
 \end{aligned}
\end{align*}
and the robot's dynamics \eqref{eq:system} are restated as
\begin{align}\label{der2}
 m_{q(t)}(c_i''(s,\theta)\dot{s}^2+c_i'(s,\theta)\ddot{s})=u_i, i\in\{1,2\}.
\end{align}
With the employed arc-length parameterization, the robot traverses the curve at constant speed, i.e., $\|c'(s,\theta))\|=\alpha$, where $\alpha$ is the arc-length of $c$. Substituting $u=u_c(t)[\cos{(\varphi)}, \sin{(\varphi)}]^T$ in polar coordinates in \eqref{der2} and using $\|\ddot{c}(s,\theta)\|=\|u\|/m_{q(t)}$, \eqref{eq:system} is equivalently restated by \eqref{parameterization} and the state $\bar{x}=[s\text{ } \dot{s}]^T$ with dynamics
\begin{align}\label{eq:paramdyn}
\begin{aligned}
\dot{\bar{x}}(t){=}&\bar{f}_q(\bar{x}(t),u_c(t),\theta)\\
=&\begin{bmatrix}\bar{x}_2&
            \sqrt{\frac{u^2_c(t)}{\alpha^2 m^2_{q(t)}}-\frac{\left(c_1'' c_2'-c_1' c_2''\right)^2\bar{x}_2^{4}}{\alpha^4}}-\frac{(c_1''c_1'+c_2''c_2')\bar{x}_2^2}{\alpha^2}\end{bmatrix}^T.
            \end{aligned}
\end{align}
To simplify the analysis in the following, the necessary optimality conditions for $u_c(t)$ will be derived for
\begin{align}\label{eq:reduceddyn}
\dot{\bar{x}}(t)\approx\begin{bmatrix}
            0&1\\0&0
           \end{bmatrix}\bar{x}(t)+\frac{1}{\alpha m_{q(t)}}\begin{bmatrix}
           0\\1\end{bmatrix}u_c(t),
\end{align}
which represents a reasonable approximation of \eqref{eq:paramdyn} along general Fourier series curves. Note that, for lines, $c_1''=c_2''=0$ implies $c_1'' c_2'-c_1' c_2''=0$ and $c_1'' c_1'+c_2' c_2''=0$, and \eqref{eq:reduceddyn} describes the dynamics of the robot exactly. Since the sensor footprint \eqref{sightreg} is typically much smaller than $\mathcal{Y}_g$, for evaluating the cost-to-go we assume that prior to their discovery all objects are located on \eqref{parameterization}, i.e., $\forall p^{(l)}\in p, p^{(l)}=c(s_l,\theta), s_l\in[0,1]$, and neglect the sensing range of the robot. A preliminary version of this analysis was presented in \cite{Nenchev2014}. In what follows, we further assume that the high-level OCP (presented in the following section) provides an optimal parameter $\theta^*$, such that the robot moving along \eqref{parameterization} with $\theta^*$ plans to cover the remaining space $\tilde{\mathcal{Y}}(t)$ as long as there are objects to be detected, and passes through object 
locations that have been discovered previously but have not been picked-up yet. We start the analysis assuming that there is only one object, i.e. $O=\{o\}$, with mass $m$ located at $\bar{x}_1\in[0,1]$. 

\subsubsection{Optimal control for one object}

The robot with dynamics \eqref{eq:reduceddyn} starts at $\bar{x}(0)=[0\text{ }0]^T$, $q(0)=(\emptyset,\emptyset,\emptyset)$ and $m_{q(0)}=m_\emptyset$. Clearly, the optimal control solving Problem~1 is divided into three parts, i.e. $u_c=(u_c^1,u_c^2,u_c^3)$, denoting the control until detection, the control until pick-up and the control until drop-off. After the object is detected at time $t_1$, when the robot moves with velocity  $\bar{x}_2(t_1)\geq 0$, it can be reached at time $t_2$ with $\bar{x}_2(t_2)=0$ by employing a time-optimal bang-bang controller with a switching at time $\tilde{t}_1$ \cite{Bryson1975}, i.e.,
\begin{align*}
u_{c}^2(t)=\begin{cases}
-1,&t\in[t_1, \tilde{t}_1),\\
1,&t\in[\tilde{t}_1,t_2).
\end{cases}
\end{align*}
Solving \eqref{eq:reduceddyn} with $u^2_{c}$ and $m_{q(0)}=m_{\emptyset}$, and applying the boundary conditions for the object's pick-up $\bar{x}_1(t_1)=\bar{x}_1(t_2), \bar{x}_2(t_2)=0$, yields the optimal cost
\begin{align}\label{cost21}
(t_2-t_1)=\underbrace{(1+\sqrt{2})}_{c_g}\alpha m_{\emptyset}\bar{x}_2(t_1).
\end{align}
Since the robot stops at $t_2$, steering it back to the depot by $u^3_{c}$ is again given by bang-bang control \cite{Bryson1975}. Since its corresponding cost is independent of $\bar{x}(t)$, $t\in[0,t_2)$, it can be neglected for finding $u^1_{c}$. 

In the worst case, the object is located the furthest away from the initial point, i.e., at $s=1$. Thus, the time-optimal control satisfies the condition
\begin{align}\label{wcconstr}
u_{c,w}^{1*}(\bar{x})=\begin{cases}
              1, &\text{if }\bar{x}_1\in[0,0.5),\\
              -1, &\text{if }\bar{x}_1\in[0.5,1].
             \end{cases}
\end{align}
In the probabilistic case, the object's location is uniformly distributed over $[0,1]$. To compute $u_{c,p}^{1*}$ we need to consider the time from detection to pick-up \eqref{cost21}, yielding $t_2=t_1+c_g\alpha  m_{\emptyset}\bar{x}_2$. To obtain a standard representation for the cost, introduce an additional state for the unknown detection time $\tilde{x}_3=t_1$, leading to an extended system state $\tilde{x}=[\bar{x}^T\text{  } t_1]^T$ with dynamics \eqref{eq:reduceddyn} and $\dot{\tilde{x}}_3=1$. Substituting the relation $d\tilde{x}_1=\tilde{x}_2 dt$, the expected time for picking up the object is
\begin{align*}
E\{t_2\}=&E\{\tilde{x}_3+c_g \alpha m_{\emptyset}\tilde{x}_2\}\\
=&\int_{0}^{1} (\tilde{x}_3+c_g m_{\emptyset}\alpha \tilde{x}_2) d\tilde{x}_1\\
=&\int_{0}^{t_2} (\tilde{x}_2 \tilde{x}_3+c_g\alpha m_{\emptyset}\tilde{x}_2^2)dt,
\end{align*}
where $t_2$ is free and the boundary constraints
\begin{align*}
\tilde{x}(0)=[0 \text{ } 0 \text{ } 0]^T,\text{ } \tilde{x}_1(t_2)=1, \text{ }\tilde{x}_2(t_2)=0
\end{align*}
must be satisfied. Thus, the probabilistic OCP has been transformed into a free final time nonlinear OCP. The corresponding control Hamiltonian is
\begin{align}\label{eq:hamil}
\begin{aligned}
 &H(\tilde{x},\lambda, u_{p,1})=(\tilde{x}_2\tilde{x}_3{+}c_g\alpha m_{\emptyset}\tilde{x}^2_2){+}\lambda^T(t)\left[ \begin{array}{ccc}\tilde{x}_2 & \frac{u_{c,p}^1}{m_\emptyset\alpha}& 1 \end{array}\right]^T,
 \end{aligned}
\end{align}
with absolutely continuous costate dynamics
\begin{align}\label{eq:adj1}
\begin{aligned}
 &\dot{\lambda}(t){=}-\frac{\partial H}{\partial \tilde{x}}=-\left[
 \begin{array}{ccc}
0 & 2c_g\alpha m_{\emptyset}\tilde{x}_2+\tilde{x}_3+\lambda_1 & \tilde{x}_2
\end{array}\right]^T.
               \end{aligned}
\end{align}
Applying Pontryagin's Minimum Principle, there exists an optimal state $\tilde{x}^*$, a control $u_{c,p}^{1*}$, and a nontrivial costate $\lambda^*$ trajectory, such that $\forall t \in [0,t_2)$,
\begin{align}\label{eq:hamilt}
H(\tilde{x}^*,\lambda^*,u_{c,p}^{1*})\leq H(\tilde{x}^*(t),\lambda(t)^*,u_{c,p}^{1}(t)),
\end{align}
leading to the following theorems.
\begin{theorem}
The optimal control is $u_{c,p}^{1*}(t)\in\left\{-1,-\frac{1}{2c_g},1\right\}$ for almost all $t\in[0,t_2)$.
\end{theorem}

\begin{theorem}\label{lem3}
  The optimal control for $\tilde{x}_1\in[0,1)$ is
 \begin{align}\label{probconstr}
  u_{c,p}^{1*}(\tilde{x})=\begin{cases}
        1, &\text{if }\tilde{x}_1\in[0,\frac{1}{3+2\sqrt{2}}),\\
        -\frac{1}{2+2\sqrt{2}}, &\text{if }\tilde{x}_1\in[\frac{1}{3+2\sqrt{2}},1].
       \end{cases}
 \end{align}
\end{theorem}
The proofs can be found in Appendix~C.

\subsubsection{Control for multiple objects}

\begin{figure}
\centering
\subfigure[]{
\includegraphics{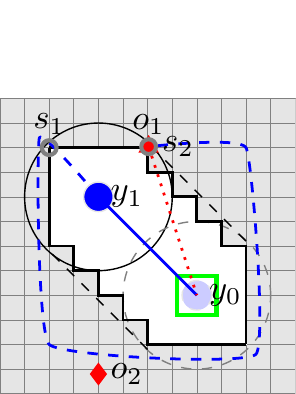}
 \label{fig:dis}
}
\subfigure[]{
\includegraphics{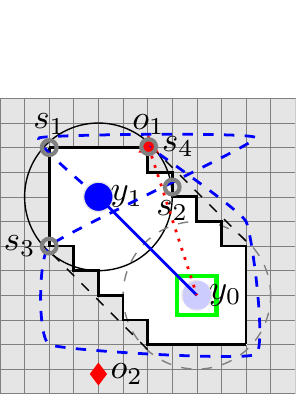}
 \label{fig:dis1}
}
\subfigure[]{
\includegraphics{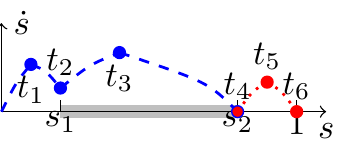}
 \label{fig:dis2}
}
\subfigure[]{
\includegraphics{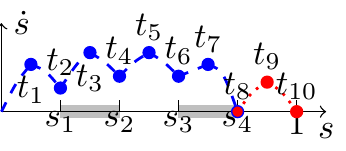}
 \label{fig:dis3}
}
\caption{Scenario upon detecting object $o_1$, denoted by $\blacktriangle$, and object $o_2$, denoted by $\blacklozenge$, with yet unknown position. Two planned position trajectories for the robot that has previously moved from $y_0=\mathbf{0}_2$ to $y_1\neq \mathbf{0}_2$ are shown with dashed lines in (a) and (b). The corresponding generalized trajectory $\bar{x}|_{[t,t_f]}$ is shown in (c) and (d), respectively.}
\label{fig:diss}
\end{figure}

For multi-object setups, we propose an approach to obtain the control analogously to the single-object case. While the worst-case optimal control is built from a finite sequence of bang-bang control segments, a computationally tractable scheme for the probabilistic control is derived as follows. Recall the robot moving in the position space shown in Fig.\,\ref{fig:discr} (b) and consider a scenario with two objects. Like in the single-object case, the robot starts moving in the discrete state $q=(\emptyset,\emptyset,\emptyset)$. The possible discrete event strings until the robot comes to a halt for the first time consist of detecting an object followed by its immediate pick-up, i.e. $\delta_i \pi_i$, or detecting both objects and stopping at one of the two objects' positions, i.e. $\delta_i\delta_j\pi_k$, $i,k,j\in\{1,2\}, i\neq j$ (see Fig.\,\ref{fig:graph}). For both cases, we employ the control \eqref{probconstr} for the time interval up to the detection of the first object, although for more than one 
object this policy may not be optimal. Now assume that object $o_1$ has just been detected. Fig.\,\ref{fig:dis} and \ref{fig:dis1} show two possible curves that provide complete exploration of $\tilde{\mathcal{Y}}(t)$, allow for picking up $o_1$ and end at the depot. The trajectories $\bar{x}^*$ resulting from the probabilistic control are shown in Fig.\,\ref{fig:dis2} and \ref{fig:dis3}, respectively. The particular control can be derived using the corresponding boundary and continuity conditions, as shown in Appendix~D. The analysis for these two cases can be easily generalized to obtain the probabilistic control for an arbitrary finite number of unexplored segments in the interval $[0,1]$.

Moving on to the problem with more than two objects, we obtain the controls along the curve by following a similar line of argumentation -- both the worst-case and the probabilistic controls consist of a finite sequence of appropriate bang-bang control segments. For that, consider the set of discrete state strings $\Sigma_q$ from a state $q(t)=q$, where all objects have been detected, to the final discrete state $q_f$, as defined in \eqref{stringset}. Then, let $\Sigma_q|_{q_3\setminus (q_1\cup q_2)}$ denote the projection of all $\sigma\in\Sigma_q$ onto $q_3\setminus (q_1\cup q_2)$. For a given $\theta$ and with $t=t_0$, the curve \eqref{parameterization} yields a discrete state string $\sigma\in\Sigma_q|_{q_3\setminus (q_1\cup q_2)}$ that is traversed in time
\begin{align}\label{timetraj}
 J_1{=}\sum_{i=1}^{d}(t_{i}{-}t_{i{-}1}){=}\sum_{i=1}^{d}\int_{\bar{x}_1(t_{i{-}1})}^{\bar{x}_1(t_{i})}\frac{1}{\bar{x}_2}d\bar{x}_1.
\end{align}
Since minimizing \eqref{timetraj} over the free parameters of the proposed policy, i.e. the switching times, can be decoupled at pick-up and drop-off instants, we solve $d-1$ Two-Point Boundary Value Problems (TPBVP's)
\begin{align}\label{lowlevelocp}
\begin{aligned}
\bar{x}^*|_{[t_{i-1},t_{i}]}=\arg &\min_{\bar{x}_1|_{[t_{i{-}1},t_{i}]}} \text{ }\int_{\bar{x}_1(t_{i{-}1})}^{\bar{x}_1(t_{i})}\frac{1}{\bar{x}_2}d\bar{x}_1,\\ \text{s.t. }& \eqref{eq:paramdyn},\bar{x}_2(t_{i-1})=\bar{x}_2(t_{i})=0,
 \end{aligned}
\end{align}
using the corresponding control trajectory $u_c$ for the worst-case or the probabilistic problem. A segment $\bar{x}^*|_{[t_{i-1},t_{i}]}$ of $\bar{x}^*=(\bar{x}^*|_{[t_0,t_{1}]},\ldots,\bar{x}^*|_{[t_{d-1},t_{d}]})$ is obtained by solving an initial value problem backward and forward in time, and choosing the minimal velocity $\bar{x}_2$ along the curve. Clearly, once all objects are detected, both versions A) and B) of Problem~1 become deterministic and can be easily solved. 

\begin{remark}\label{otherparam}
The TPBVP \eqref{lowlevelocp} can be easily extended for general convex input constraints $u\in U$ or nonlinear robot dynamics, as shown in \cite{Verscheure2009}. If the curve is not continuously differentiable, it can be approximated by a finite number of cubic spline segments, such that the OCPs remain numerically feasible. 
\end{remark}

Now the only remaining task is computing $\theta^*$ that solves the high-level optimization problem. 

\subsection{Parametric optimization}

The solution $\theta^*$ of the high-level optimization problem yields a curve \eqref{parameterization} that can be traversed optimally according to the policy outlined above, while providing complete exploration of the remaining space $\tilde{\mathcal{Y}}(t)$, passing through all positions of objects in $q_3\setminus(q_1\cup q_2)$ and ending at the depot. Let these constraints be captured by the mapping $h:\bar{x}|_{[t,t_f]} \times \mathbb{R}^a \to \mathbb{R}^b$, where $h(\bar{x}^*,\theta)=\mathbf{0}_b$, if the constraints are satisfied, and $h(\bar{x}^*,\theta)\neq\mathbf{0}_b$, else. Thus, with \eqref{timetraj} and \eqref{lowlevelocp}, the high-level optimization problem at $(q(t),(\bar{x}(t),\mathcal{W}(t)))$ reads
\begin{align}\label{costpar}
\begin{aligned}
(t_f-t)^*{=}\min_{\theta} &\text{ } J_1 \\
\text{s.t. } &h(\bar{x}^*,\theta)=\mathbf{0}_b.
\end{aligned}
\end{align}
Leveraging ideas from \cite{Gol2014,Nenchev2016}, let $J_2: \bar{x}|_{[t,t_f]} \times \mathbb{R}^a \to \mathbb{R}^b$ be a continuously differentiable version of the mapping $h$ that captures the constraints of the high-level OCP, such that $J_2(\bar{x}^*,\theta)= \mathbf{0}_b$, if $h(\bar{x}^*,\theta)=\mathbf{0}_b$, and $J_2(\bar{x}^*,\theta)> \mathbf{0}_b$, else. We will now introduce the individual constraints of the optimization problem.

To guarantee that all objects are detected eventually, the robot has to observe every $w_{i_k}\in \mathcal{W}(t)\subset W, i_k\in\{1,\ldots,K\}$, when it moves along the curve. For every $w_{i_k}$, we define a function $d_k(s,\theta)$, such that $d_k(s,\theta)=0$, if $w_k$ is within the sensing range \eqref{sightreg} of the robot with position $y(t)=c(s,\theta)$, and $d_k(s,\theta)>0$, otherwise. The choice of $d_k(s,\theta)$ is not unique. In particular, we employ
\begin{align}\label{convsightreg}
d_k(s,\theta){=}\begin{cases}
1{-}\exp{(-(D_k(s,\theta){-}r)^2)}, &\text{if } D_k(s,\theta){>}r,\\
0,&\text{else,}
\end{cases}
\end{align}
where $D_k(s,\theta)=\|c(s,\theta){-}w_k\|$. This leads to the constraint vector $\mathcal{D}_\mathcal{W}=[\min_{s\in[0,1]}d_1(s,\theta),\ldots,\min_{s\in[0,1]}d_{\tilde{K}}(s,\theta)]$, $\tilde{K}=|\mathcal{W}(t)|$, which is required to be equal to $\mathbf{0}_{1,\tilde{K}}$, as every $w_{i_k}\in\mathcal{W}(t)$ has to be seen along the curve.

In addition, all objects that have been detected, but have not been picked up or dropped at the depot yet, i.e. all $o_l\in q_3\setminus(q_1\cup q_2)$, have to be eventually picked up, while the robot moves along the curve. Since the requirement that the robot  performs a pick-up with zero velocity is taken care of by the low-level OCP, in the high-level OCP we only have to guarantee that the curve passes through all corresponding positions $p^{(l)}$. Hence, for every $p^{(i_l)}, i_l\in \{1,\ldots,\tilde{L}\}, \tilde{L}=|q_3\setminus(q_1\cup q_2)|$, we define a function $\tilde{d}_l(s,\theta)$, such that $\tilde{d}_l(s,\theta)=0$, if $p^{(i_l)}=c(s,\theta)$, and $\tilde{d}_l(s,\theta)>0$, otherwise. In particular, we employ $\tilde{d}_l(s,\theta)=1{-}\exp{(-\|c(s,\theta)-p^{(i_l)}\|^2)}$. Note that the choice of $\tilde{d}_l$ is not unique. This leads to the constraint vector $\mathcal{D}_{q_3{\setminus}(q_2{\cup}q_1)}=[\min_{s\in[0,1]}\tilde{d}_1(s,\theta), \ldots, \min_{s\in[0,1]}\tilde{d}_{\tilde{L}}(s,\
theta)]$, which is required to be equal to $\mathbf{0}_{1,\tilde{L}}$, since every $p^{(i_l)}$ has to be visited at some point along the curve.

Further, upon a detection at $s_-=s(t)$, if the robot's velocity is not zero, the robot is required to continue its motion in a smooth manner, despite changing from moving along the curve characterized by the previous parameter vector denoted by $\theta_{-}$, to moving along the curve characterized by the current parameter vector denoted by $\theta$. Hence, we introduce a function $d^y_{in}(s_-,\theta_-,\theta)$, such that $d^y_{in}(s_-,\theta_-,\theta)=0$ when $c(s_-, \theta_{-})=c(0,\theta)$, and $d^y_{in}(s_-,\theta_-,\theta)>0$, else. In particular, we employ $d^y_{in}(s_-,\theta_-,\theta)=1{-}\exp{(-\|c(s_-,\theta_{-})-c(0,\theta)\|^2)}$. Analogously, introduce a function $d^v_{in}(s_-,\theta_-,\theta)$, such that $d^v_{in}(s_-,\theta_-,\theta)=0$ when $c'(s_-,\theta_{-})=c'(0,\theta)$, and $d^v_{in}(s_-,\theta_-,\theta)>0$, else. In particular, we employ $d^y_{in}(s_-,\theta_-,\theta)=1{-}\exp{(-\|c'(s_-,\theta_{-})-c'(0,\theta)\|^2)}$. Note that both $d^y_{in}(s_-,\theta_-,\theta)$ and $d^v_{in}(s_-, \
theta_-,\theta)$ are not unique. Finally, the robot is required to return to the depot in order to drop all objects. Since the requirement that the robot performs a drop-off with zero velocity is taken care of by the low-level OCP, in the high-level OCP we just have to guarantee that the curve ends at the depot. Thus, we introduce the function $d(\theta)$, such that $d(\theta)=0$, if $c(1,\theta)=y_d$, and $d(1,\theta)>0$, otherwise. In particular, we use $d(\theta)=1{-}\exp{(-\|c(1,\theta){-}y_d\|^2)}$. Note that the choice of $d(\theta)$ is not unique. This leads to the constraint vector
\begin{align*}
 \mathcal{D}_{in,f}=\begin{cases}
                     [d^y_{in},d^v_{in},d], &\text{if }\bar{x}_2\neq0,\\
                     [d^y_{in},d], &\text{else},
                    \end{cases}
\end{align*}
which is required to be equal to zero. Once all objects are discovered, exploration is no longer necessary and the constraints $\mathcal{D}_\mathcal{W}$ are neglected. Thus, the constraints $J_2$ are given by
\begin{align}\label{uncertainty}
 J_2(\bar{x}^*,\theta){=}\begin{cases}
      [\mathcal{D}_\mathcal{W}, \mathcal{D}_{q_3{\setminus}(q_2{\cup}q_1)}, \mathcal{D}_{in,f}]^T,&\text{if }q_3(t){\neq}O,\\
      [\mathcal{D}_{q_3{\setminus}(q_2{\cup}q_1)}, \mathcal{D}_{in,f}]^T,&\text{else}.
     \end{cases}
\end{align}

The constrained optimization problem \eqref{costpar} is approximately solved with \eqref{uncertainty} by the augmented Lagrangian method \cite{Bertsekas1996}, yielding the iterative unconstrained optimization problem
\begin{align}\label{costp}
\theta^*_z=\arg \min_{\theta} \hat{J} =J_1+\frac{\mu_z}{2} J_2^T J_2+\bar{\lambda}_z^T J_2,
\end{align}
where $\mu_z\in\mathbb{R}_{>0}$ is an optimization tuning variable that increases with each iteration and $\bar{\lambda}_z\in\mathbb{R}^b$ is an estimate of the Lagrangian multiplier, updated by 
\begin{align}\label{lambdak}
\bar{\lambda}_{z+1}=\bar{\lambda}_z+\mu_z J_2(\bar{x}^*,\theta_z^*) 
\end{align}
for every iteration $z=0,1,2,\ldots$ 

Since \eqref{uncertainty} is continuously differentiable w.r.t. $s$ and $\theta$, the unconstrained optimization problem \eqref{costp} can be solved by gradient-based optimization. Taking into account the dynamics of the hybrid automaton $\mathcal{H}$ and substituting $\alpha=\|c'(\bar{x}_1,\theta)\|$ for the employed arc-length parameterization, the gradient of \eqref{costp} (omitting function arguments) is given by
\begin{align}\label{gradj1}
\begin{aligned}
 \nabla_\theta \hat{J}{=}&\sum_{n=1}^{N} \nabla_\theta \int_{t_{n{-}1}}^{t_{n}}dt{+}\sum_{\beta=1}^b (\mu_z J_{2,\beta}{+}\bar{\lambda}_{z,\beta}) \nabla_\theta J_{2,\beta}\\
=&\sum_{n=1}^{N}\nabla_\theta  \int_{t_{n{-}1}}^{t_{n}}\hspace{-0.5em}\frac{\|c'\|}{\alpha} dt{+}\sum_{\beta=1}^b (\mu_z J_{2,\beta}{+}\bar{\lambda}_{z,\beta}) \nabla_\theta J_{2,\beta}.
\end{aligned}
\end{align}
Observing that $\bar{x}_1$ depends on $\theta$ through \eqref{eq:paramdyn}, we employ Infinitesimal Perturbation Analysis (IPA) \cite{Cassandras2010a} to obtain the gradient $\nabla_\theta  \int_{t_{n{-}1}}^{t_{n}}\hspace{-0.5em}\frac{\|c'\|}{\alpha} dt$ (and, thus,  $\nabla_\theta \hat{J}$). Over an interval $\tau_n=[t_{n{-}1},t_{n})$, $n=1,\ldots,N$, the evolution of $\bar{x}$ is described by the vector field $\bar{f}_{n{-}1}(t,\bar{x},\theta)$ (with a slight abuse of notation). For $t\in[t_{n{-}1},t_{n})$,
\begin{align}\label{ipa1}
 \frac{d}{dt}\nabla_\theta \bar{x}(t,\theta){=}\frac{\partial \bar{f}_{n{-}1}(t,\bar{x},\theta)}{\partial \bar{x}} \nabla_\theta \bar{x}(t,\theta){+}\nabla_\theta \bar{f}_{n{-}1}(t,\bar{x},\theta)
 \end{align}
holds with the boundary condition
\begin{align}\label{ipa2}
\begin{aligned}
 \nabla_\theta \bar{x}(t_{n{-}1}^+,\theta){=}&\nabla_\theta \bar{x}(t_{n{-}1}^-,\theta)
 {+}\\&[\bar{f}_{{n-2}}(t_{n{-}1}^-,\bar{x},\theta){-}\bar{f}_{n{-}1}(t_{n{-}1}^+,\bar{x},\theta)]\nabla_\theta t_{n{-}1}(\theta).
 \end{aligned}
\end{align}
Thus, we obtain
\begin{align}\label{ipa3}
 \nabla_\theta \bar{x}(t,\theta){=}\nabla_\theta \bar{x}(t_{n{-}1}^+,\theta)+\int_{t_{n{-}1}}^{t} \frac{d}{dt}\nabla_\theta \bar{x}(t,\theta) dt.
\end{align}
Since $\bar{f}_{1,{n-2}}(t_{n{-}1}^-,\bar{x},\theta){=}\bar{f}_{1,n{-}1}(t_{n{-}1}^+,\bar{x},\theta)$, from \eqref{ipa2}, we observe that $\nabla_\theta \bar{x}_1(t_{n{-}1}^+,\theta)=\nabla_\theta \bar{x}_1(t_{n{-}1}^-,\theta)$. Using \eqref{ipa1} with $\frac{\partial \bar{f}_{1,n{-}1}(t,\bar{x},\theta)}{\partial \bar{x}_1}=0$ and $\nabla_\theta \bar{f}_{1,n{-}1}(t,\bar{x},\theta)=0$ and \eqref{ipa3}, the gradient \eqref{gradj1} (omitting function arguments) is obtained by
\begin{align}\label{gradj}
\begin{aligned}
 \nabla_\theta \hat{J}{=}\sum_{i=1}^{d} \int_{t_{i{-}1}}^{t_{i}}\hspace{-0.5em}\nabla_\theta \frac{\|c'(\bar{x}_1,\theta)\|}{\alpha} dt{+}\sum_{\beta=1}^b (\mu_z J_{2,\beta}{+}\bar{\lambda}_{z,\beta}) \nabla_\theta J_{2,\beta}.
\end{aligned}
\end{align}

The partial derivatives of \eqref{convsightreg} w.r.t. $s$ or $\theta$ (omitting function arguments) are obtained by 
\begin{align}  \label{dobscost} 
&\nabla_{(.)} d_k{=}\begin{cases}
2(D_k{-}r)\exp{(-(D_k{-}r)^2)}\frac{\partial D_k}{\partial (.)}, &\text{if } D_k{>}r,\\
0,&\text{else,}
\end{cases}
\end{align}
where $\frac{\partial D_k}{\partial (.)}{=}\frac{1}{D_k}\left((c_1{-}w_{1,k})\nabla_{(.)} c_1{+}(c_2{-}w_{2,k})\nabla_{(.)} c_2\right)$. The corresponding derivatives of $\tilde{d}_l$ and $d$ are obtained analogously. The required partial derivatives for Fourier series are shown in Appendix~B. Then, to compute \eqref{gradj}, we obtain the optimal solutions of the internal OCPs of the constraints $\mathcal{D}_{\mathcal{W}}$ by a gradient-based algorithm, i.e.,
\begin{align}\label{innerocp}
 s_{z+1}=s_z-\eta_z\left.\frac{\partial d_k(s_z,\theta)}{\partial s}\right|_{s\in[0,1]},
\end{align}
where $\{\eta_z\},z=0,1,2\ldots$ is an appropriate step size sequence, $\left.\frac{\partial d_k}{\partial s}\right|_{s\in[0,1]}$ is the gradient projected onto the feasible interval $s\in[0,1]$ and the algorithm terminates when $\left|\left.\frac{\partial d_k}{\partial s}\right|_{s\in[0,1]}\right|<\epsilon$ (for a given threshold $\epsilon$). Analogously, we solve the internal OCPs of the constraints $\mathcal{D}_{q_3\setminus(q_1\cup q_2)}$, yielding the value of $J_2$. Then, $\nabla_\theta J_2$ is computed by \eqref{dobscost} for the (local) optima acquired by solving \eqref{innerocp} for the corresponding constraints.

As the considered setup is static, the gradient obtained through IPA is a trivially unbiased estimate of the gradient of the cost for Problem~B. Thus, the optimal parameter vector $\theta^*$ is obtained with a gradient-based algorithm, i.e.,
\begin{align}\label{gradsolve}
 \theta_{z+1}=\theta_z-\eta_z \left. \nabla_\theta \hat{J}(\bar{x}^*,\theta_z)\right|_{\mathcal{Y}_g}
\end{align}
where $\{\eta_z\},z=0,1,2,\ldots$ is a properly selected step-size sequence and the gradient is projected onto the feasible position space $\mathcal{Y}_g$. The algorithm terminates when $|\nabla_\theta \hat{J}(\bar{x}^*,\theta_z)|_{\mathcal{Y}_g}|<\epsilon$ for a pre-specified threshold $\epsilon$. Note that the OCPs \eqref{innerocp} are non-convex, the solution of \eqref{costp} acquired by \eqref{gradsolve} will, in general, be only locally optimal.

\begin{algorithm}[t]
\caption{Event-driven receding horizon control upon detection}\label{alg}
  \begin{algorithmic}[1]
    \REQUIRE Robot dynamics described by the hybrid automaton $\mathcal{H}$ with $\text{Init}=(q(t),(\bar{x}(t),\mathcal{W}(t)))$; curve $y(t)=c(\bar{x}_1,\theta)$ with initial parameter vector $\theta$; optimization parameters $0<\epsilon\ll1$, $\nu>1$
    \ENSURE The optimal control $u|^*_{[t,t_f]}$
    \STATE Set $\mu_1=1$ and $\bar{\lambda}_1=\mathbf{0}_b$.
    \WHILE{$J_2(\bar{x}^*,\theta)>\epsilon \mathbf{1}_b$}
         \REPEAT 
         \STATE Compute $u^*(\bar{x}^*)$ through \eqref{lowlevelocp} and $J_2$ and $\nabla_\theta J_2$ through \eqref{innerocp} for $\theta$ for the worst-case or the probabilistic case.
	  \STATE Compute $\hat{J}$ and $\nabla_\theta \hat{J}|_{\mathcal{Y}_g}$ with $\mu_z$, and update $\theta$ through \eqref{gradsolve} and $\bar{\lambda}_z$ through \eqref{lambdak}.
	 \UNTIL{$|\nabla_\theta \hat{J}|_{\mathcal{Y}_g}|<\epsilon$}
	 \STATE Set $\mu_{z+1}=\nu\mu_z$.
 \ENDWHILE
	\RETURN $u|^*_{[t,t_f]}=u^*(\bar{x}^*)$
		\end{algorithmic}
\end{algorithm}

The overall event-driven solution is summarized in Alg.\,\ref{alg}. Upon a detection of an object, as long as the current solution violates the constraints, \eqref{lowlevelocp} and \eqref{gradsolve} are solved iteratively, where $\mu_z$ increases with each iteration, thus increasing the importance of $J_2$ over $J_1$ in the optimization. A detection yields an additional constraint for the high-level OCP, which will be violated initially, in general. Assuming that all other constraints were satisfied in the previous run, the algorithm typically terminates within a small number of iterations.

Before the first detection, a good initial guess for $\theta$ is required, such that Alg.\,1 produces a good local optimum despite the non-convexity of \eqref{costp}. A good initial guess is often obtained by randomized optimization, e.g. using a Stochastic Comparison Algorithm \cite{Bao1996}.

\begin{remark}
Under the presence of an obstacle obs with a-priori unknown size and location, the obstacle is approximated by all $w_k$ of the discretization of $\mathcal{Y}_g$ that have been covered by the sensor until time $t$ and belong to the obstacle region, i.e. all $w_k\in \cup_{\tilde{t}\in[0,t]} \mathcal{O}(y(\tilde{t}))\land w_k\in \text{obs}$. Then, $J_2$ is augmented by a term that is only active in the surroundings $d_s=d_g/\sqrt{2}$ of the corresponding $w_k$, $k\in\{1,\ldots,\tilde{K}\}$, and is continuously differentiable w.r.t. $s$ or $\theta$, e.g., 
\begin{align}
 J_{o}=\sum_{k=1}^{\tilde{K}} \max\{0,1-(D_k/d_s)^2)\}^2.
\end{align}
With the partial derivative
\begin{align*}
 \nabla_{(.)} J_{o}{=}&{-}\frac{4}{d_s^2}\sum_{k=1}^{\tilde{K}}\left[\frac{\partial D_k}{\partial (.)}\max\{0,1{-}\frac{D^2_k}{d_s^2}\}\right],
\end{align*}
the (local) minimum of $J_o^*$ is acquired by solving an OCP of the form \eqref{innerocp}, and the corresponding $\nabla_{\theta} J_{o}$ is used to solve \eqref{gradsolve}. Note that $J_o=0$ and $\nabla_\theta J_o=\mathbf{0}_a$, if the curve does not intersect the obstacle. Alg.\,\ref{alg} remains unchanged. In general, the non-convexity of the optimization space increases with a growing number of obstacles, which may deteriorate the quality of the local minima obtained by Alg.\,1. In environments with high obstacle densities, one should choose parameterizations with a higher degree of freedom (i.e. larger $\Gamma_1$ and $\Gamma_2$ in the case of Fourier series) to allow for efficient obstacle avoidance. 
\end{remark}

\subsection{Complexity}

The time-driven approaches are computationally expensive due to their NP-hardness and the generally increasing non-convexity of $\tilde{\mathcal{Y}}(t)$ during mission execution. The number of possible discrete state strings $|\Sigma_q|$ is exponential in the number of not yet detected objects. In our implementation, the time-driven approaches remained computationally feasible for $|L|\leq 4$. In contrast, the event-driven approaches rely on solving a finite number of OCPs by gradient-based methods with significantly lower computational cost, as indicated by the numerical example in the following.

\section{Numerical example}\label{sec:example}

\subsubsection{Implementation}
The methods were implemented in MATLAB, using the solver IPOPT for the MIPs of the time-driven and $ode45$ for the event-driven methods. All computations were performed on an Intel\textsuperscript{\textregistered} Core\textsuperscript{\texttrademark} i7 2.20 GHz processor with 8 GB RAM.

\subsubsection{Setup}
Consider the bounded position space $\mathcal{Y}_g=[-5,5]\times [-5,5]\,\text{m}$ and its regular discretization (as described in Section~\ref{sec:solution}) with a grid constant of $0.25$\,m. Let the object set be $O=\{o_1,o_2,o_3\}$ with $p^{(1)}=[-3.1,-3.1]^T,m^{(1)}=1\,\text{kg}$, $p^{(2)}=[1.9,-1.9]^T,m^{(2)}=2\,\text{kg}$ and $p^{(3)}=[3,3]^T,m^{(3)}=2\,\text{kg}$. Initially, the positions of the objects are unknown to the robot. The robot with sensor footprint of size $r=1$\,m and nominal mass $m_{\emptyset}=2\,\text{kg}$ starts at the depot $y_d=\mathbf{0}_2$ at rest. A static obstacle with a priori unknown location and size to the robot, is described by
\begin{align*}
 \text{obs}{=}\left\{y\in \mathbb{R}^2{:}\begin{bmatrix}
\mathbf{I}_2\\
-\mathbf{I}_2
\end{bmatrix}\!y\leq\!\begin{bmatrix}
\mathbf{I}_2\\
-\mathbf{I}_2
\end{bmatrix} \begin{bmatrix}
1.875\\-3.625
\end{bmatrix}\!{+}\!\begin{bmatrix}
\mathbf{I}_2\\
\mathbf{I}_2
\end{bmatrix} \begin{bmatrix}
0.375\\0.375
\end{bmatrix}
\right\}\!.
\end{align*}
The sampling time $t_s=0.2$\,s and the optimization horizon $N_{\text{max}}=8$ were chosen for the time-driven approaches. Fourier series of order $\Gamma_1=\Gamma_2=3$ were used for the event-driven approaches. Optimization was performed with parameters $\nu=2,\epsilon=10^{-3}$, and a maximal integration time of $40$\,s for $ode45$. The parameter vector $\theta$, which characterizes the curve before the first detection, is obtained by running Alg.\,1 for $100$ random initializations and selecting the best (local) optimum.

\subsubsection{Analysis}
\begin{figure*}
 \centering
  
 
 \subfigure[$t^{(3)}=4$\,s]{
\includegraphics{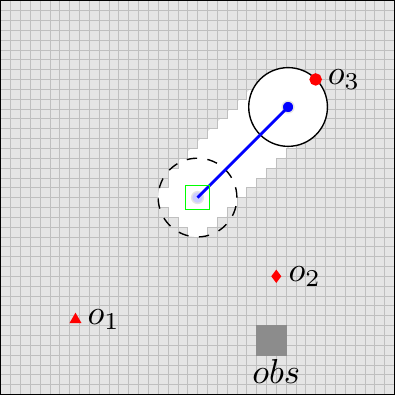}
 \label{fig:simsdwc1}
}
\subfigure[$t^{(2)}=15$\,s]{
\includegraphics{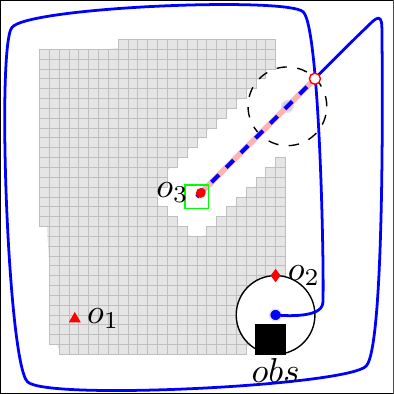}
 \label{fig:simsdwc2}
}
\subfigure[$t^{(1)}=25$\,s]{
\includegraphics{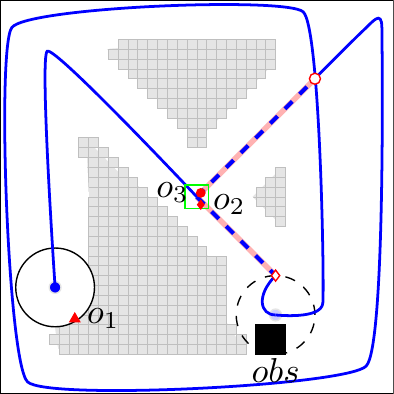}
 \label{fig:simsdwc3}
}
\subfigure[$t_f=32$\,s]{
\includegraphics{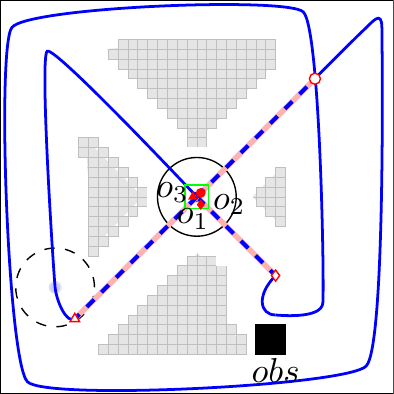}
 \label{fig:simsdwc4}
}

 \subfigure[$t^{(2)}=11.2$\,s]{
\includegraphics{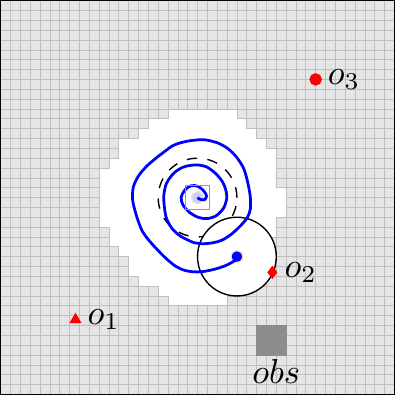}
 \label{fig:simsdpr1}
}
 \subfigure[$t^{(3)}=23.2$\,s]{
\includegraphics{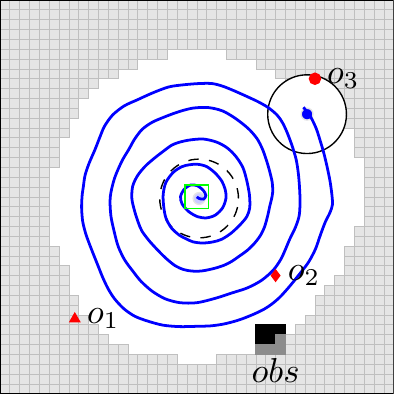}
 \label{fig:simsdpr2}
}
 \subfigure[$t^{(1)}=32.2$\,s]{
\includegraphics{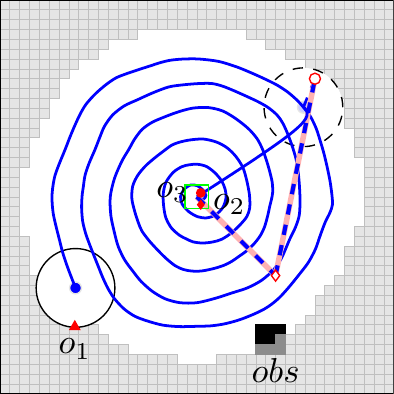}
 \label{fig:simsdpr3}
}
 \subfigure[$t_f=37.2$\,s]{
\includegraphics{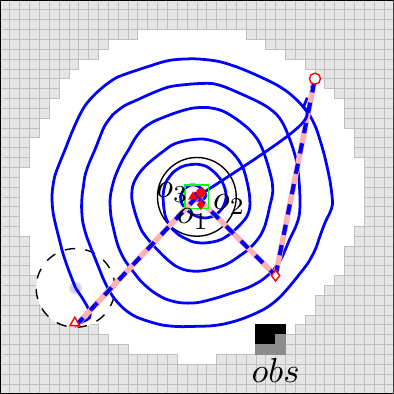}
 \label{fig:simsdpr4}
}

 
 \subfigure[$t^{(3)}=3.8$\,s]{
\includegraphics{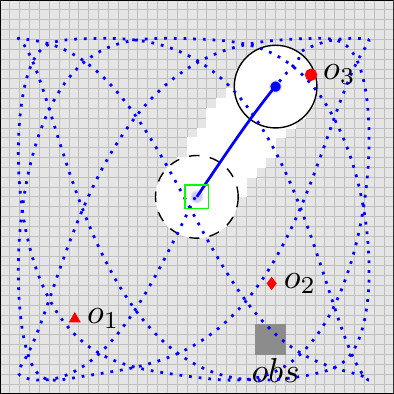}
 \label{fig:simparwc1}
}
 \subfigure[$t^{(2)}=9$\,s]{
\includegraphics{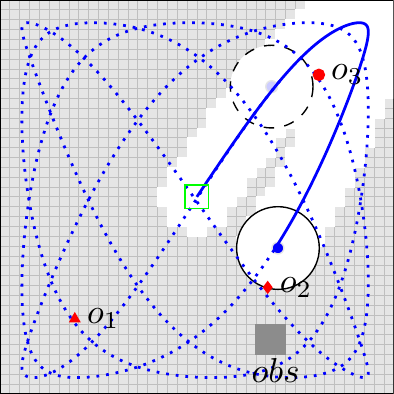}
 \label{fig:simparwc2}
}
 \subfigure[$t^{(1)}=30.2$\,s]{
\includegraphics{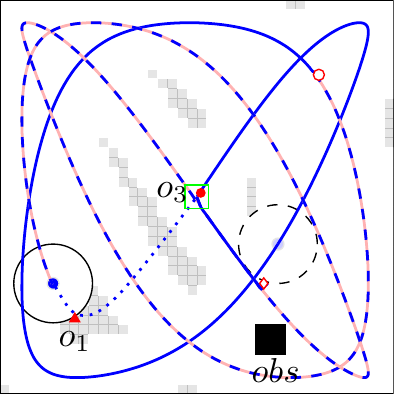}
   \label{fig:simparwc3}
}
 \subfigure[$t_f=36.2$\,s]{
\includegraphics{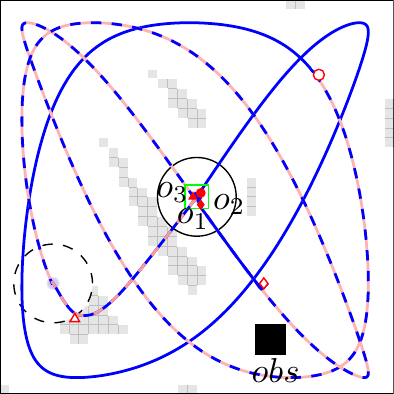}
   \label{fig:simparwc4}
}


 \subfigure[$t^{(2)}=4.4$\,s]{
\includegraphics{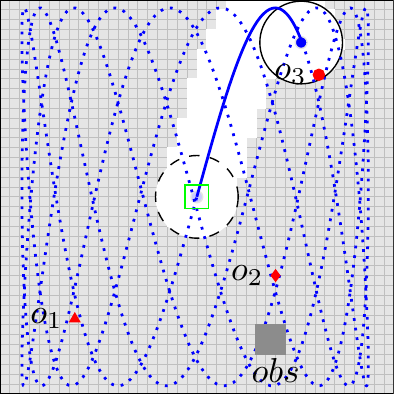}
   \label{fig:simparpr1}
}
 \subfigure[$t^{(3)}=14$\,s]{
\includegraphics{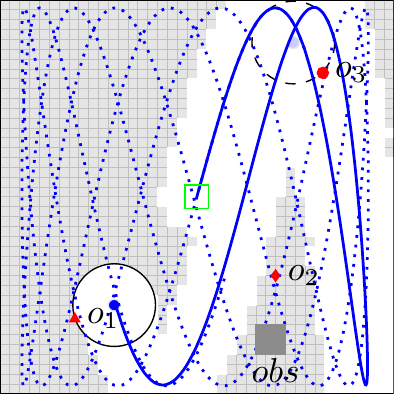}
   \label{fig:simparpr2}
}
 \subfigure[$t^{(1)}=25.4$\,s]{
\includegraphics{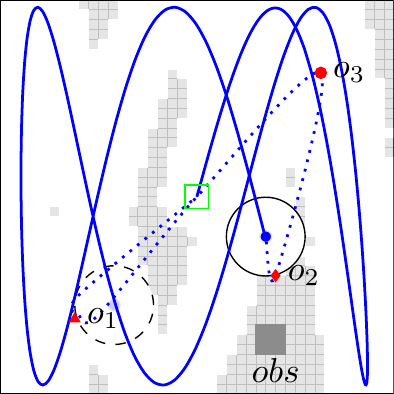}
   \label{fig:simparpr3}
}
 \subfigure[$t_f=32.2$\,s]{
\includegraphics{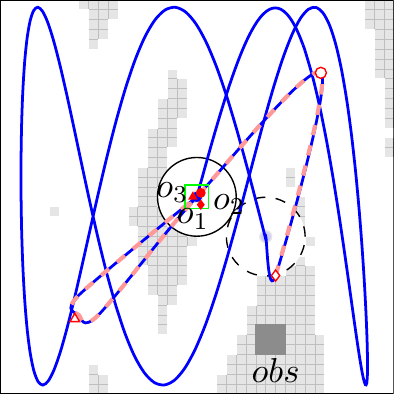}
   \label{fig:simparpr4}
}

\caption{Snapshots of the robot's motion at detection instants $t^{(l)}$ and the final time $t_f$ respectively obtained with the time-driven worst-case (a-d) and probabilistic (e-h), and the event-driven worst-case (i-l) and probabilistic (m-p) methods. Executed paths with nominal dynamics are denoted by solid, executed paths in other dynamical modes by thicker dashed, and planned trajectories by dotted lines.}
\label{fig:archcov}
\end{figure*}

Fig.\,\ref{fig:archcov} shows snapshots of the robot's motion at object detection instants and the final time, obtained by applying the presented methods. Snapshots at obstacle detection instants were omitted due to space limitations. As it can be seen in the plots, the robot successfully avoids the obstacle. The time-driven and the event-driven methods lead to qualitatively similar solutions with small performance differences. On average, online re-computation took $35$\,s and $21$\,s for the worst-case and probabilistic time-driven approaches (performed at every time instant), and $6$\,s and $5$\,s for the worst-case and probabilistic event-driven approaches (performed only upon a detection), respectively, thus indicating that the latter two are particularly suitable for real-time computation. 

For 100 random placements of the three objects in $\mathcal{Y}_g$, the average cost for solving the task by the probabilistic event-driven solution was $30.1$\,s, thus, performing better than the worst-case event-driven solution with an average cost of $32.21$\,s. In general, the worst-case solution results in a more ``cautious'' policy including intermediate pick-ups and drop-offs. In contrast, the probabilistic evaluation typically leads to a ``threshold-based'' policy, where previously detected objects are collected in one sweep after longer exploration phases. Note that both the convergence speed and the quality of the outcomes of the event-driven approaches strongly depend on the chosen initial conditions and the step size selection method of the gradient optimization procedures.

\section{Conclusions}\label{sec:conclusion}

A time-optimal hybrid control problem for a robot that has to find and collect a finite number of objects located in a two-dimensional space and move them to a depot has been addressed. Two approaches have been proposed for the worst- and a probabilistic case, assuming uniform distribution of the objects over the search space -- a time-driven scheme based on a space and time discretization, and an event-driven approach based on motion parameterization. The methods were compared in a numerical example, reflecting the significant computational advantage of the event-driven methods, while yielding similar qualitative results to the time-driven ones. The worst-case evaluation of the cost typically provides a ``cautious'' solution, often resulting in an intermediate pick-up and drop-off, while the probabilistic evaluation leads to a ``threshold-based'' policy, consisting of longer exploration phases and multiple successive pick-up's with a following drop-off. Future work will address a multi-robot setup and alternative optimization techniques for the event-driven approaches, e.g. using the Alternating Direction Method of Multipliers (ADMM) \cite{Boyd2011}.

\section*{Appendix A: Constraint set}

The constraint set $\mathcal{C}(i)$ at time $i$ is constructed successively for $\mathcal{H}$ by introducing additional slack variables. We provide a brief overview of the most important aspects of the implementation based on examples. Let $i,j\in[0,N_{\text{max}}-1]$ be time instants in the optimization horizon. A depot visit at $i$ is captured by a Boolean vector $b_d$ with
\begin{align*}
 b_{d_i}=\begin{cases}
        1, &\text{if } x_i=[y_d^T\text{ }\mathbf{0}_2^T]^T \land b_{d_j}=0, \forall j<i,\\
        0, &\text{else.}
        \end{cases}
\end{align*}
Then, a switching of the time-discretized dynamics (obtained by equidistant sampling of \eqref{eq:system}) from $q_1$ to $q_\emptyset$ is realized by
\begin{equation*}
\forall i, x_{i+1}=\begin{cases}
\tilde{f}_{q_1}(x_i,u_i), &\text{if } b_{d_j}=0,\forall j\leq i,\\
\tilde{f}_{q_\emptyset}(x_i,u_i), &\text{else,}
\end{cases}
\end{equation*}
implying that the system evolves in state $q_1$ until $b_{d_j}=1$, when it switches to and remains in $q_\emptyset$ thereafter. The input constraints are approximated by a regular $Z$-sided polygon, i.e., 
\begin{align*}
 \forall \in \{1,\ldots,Z\},\forall i, u_{i,1} \sin{\left(\frac{2\pi z}{Z}\right)}{+}u_{i,2}\cos{\left(\frac{2\pi z}{Z}\right)}\leq 1.
\end{align*}
The final time constraint can be captured by a boolean vector
\begin{align*}
&\forall i, b_{f_i}{=}\begin{cases}
        1, &\text{if } x_i{=}[y^T\text{ } \mathbf{0}_2^T]^T\land y{\in}\{p^*,y_d\},\\
        0, &\text{else,}
        \end{cases}, \sum_{i=0}^{N_{\text{max}}-1} b_{f_i}{=}1,
\end{align*}
reflecting that the trajectory may end at an expected position $p^*$ of an object or at the depot. The constraints are implemented by big-$M$ relaxations, commonly employed in MIP, where the relaxation coefficients are chosen according to the setup. Finally, the cost is $\ell= t_s\sum_{i=1}^{N_{\text{max}-1}}i b_{f_i}$.

\section*{Appendix B: Fourier series trajectories}

In the event-driven approaches, we parameterize the position of the robot by Fourier series of respective order $\Gamma_1$ and $\Gamma_2$, i.e., 
\begin{align*}
 y(t)=\begin{bmatrix}
       c_1\\c_2
      \end{bmatrix}
=\begin{bmatrix}
       a^1_0+\sum_{\gamma=1}^{\Gamma_1} a^1_\gamma\sin{(4\pi^2\gamma f_1 s(t)+\phi_\gamma^1)}\\ 
       a^2_0+\sum_{\gamma=1}^{\Gamma_2} a^2_\gamma\sin{(4\pi^2\gamma f_2 s(t)+\phi_\gamma^2)}
      \end{bmatrix},
\end{align*}
where $f_1$ and $f_2$ are base frequencies, $a^1_0$ and $a_0^2$ are zero frequency components, $a^1_\gamma$ and $a_\gamma^2$ are amplitudes for the sinusoid functions with frequency $\gamma f_1$ and $\gamma f_2$, and $\phi_\gamma^1$ and $\phi_\gamma^2$ are phase differences with respect to the $(\gamma+1)$-th term of $y_1$ or $y_2$. Since only the ratio of $f_1$ and $f_2$ (and not their absolute values) determines the shape of the trajectories, $f_1$ is treated as a free parameter, while $f_2$ is kept constant. With $A_1=\begin{bmatrix}                                                                                                                                                                                                                                                                                                                                                                                                           
                                    
                       a_0^1,\ldots,a^1_{\Gamma_1}                                                                                                                                                                                                                                                                                                                                                                                                                                                              
                                                                                                               \end{bmatrix}$, $A_2=\begin{bmatrix}                                                                                                                                                                                                                                                                                                                                                                                                                                                                                                                                                                                                               a_0^2,\ldots,a^2_{\Gamma_2}                                                                                                                                                                                                                                                                 
     
                                                                                                                                                                                                                                                                                                       \end{bmatrix}$, $\Phi_1=\begin{bmatrix}                                                                                                                                                                                                                                                                                                                                                                                                                                                                                                                                                                                                               \phi_1^1,\ldots,\phi^1_{\Gamma_1}                                                                
     
                                                                                                                                                                                                                                                                                                                                                                                                                                                                                                       \end{bmatrix}$ and $\Phi_2=\begin{bmatrix}                                                                                                                                                                                                                                                                                                                                                                                                                                                                                                             
                                                                                                  \phi_1^2,\ldots,\phi^2_{\Gamma_2}                                                                                                                                                                                                                                                                                                                                                                                                                                                                                                                                                                           \end{bmatrix}$, the overall parameter vector is $\theta{=}[f_1,A_1,A_2,\Phi_1,\Phi_2]^T$. The derivative of the curve w.r.t. $\theta$ with $i\in\{1,2\}$ is
\[
 \nabla_\theta y_i=\nabla_\theta c_i(s,\theta)=\begin{bmatrix}
                                       \frac{\partial y_i}{\partial f_1}&\frac{\partial y_i}{\partial A_1}&\frac{\partial y_i}{\partial A_2}&\frac{\partial y_i}{\partial \Phi_1}&\frac{\partial y_i}{\partial \Phi_2}
                                      \end{bmatrix}^T
\]
with $\gamma=1,\ldots,\Gamma_i$ and
\begin{align*}
 &\frac{\partial y_i}{\partial f_1}{=}\begin{cases}
4\pi^2s(t) \sum_{\gamma=1}^{\Gamma_1} a^1_\gamma\cos{(4\pi^2\gamma f_1 s(t)+\phi_\gamma^1)}, & i{=}1,\\
        0,                            & i{=}2,
                                   \end{cases}\\
                           & \frac{\partial y_i}{\partial a^j_0}{=}\begin{cases}
1, & i{=}j,\\
        0,                            & i{\neq}j,
                                   \end{cases},     
\frac{\partial y_i}{\partial a_\gamma^j}{=}\begin{cases}
\sin{(4\pi^2\gamma f_i s(t)+\phi_\gamma^i)}, & i{=}j,\\
        0,                            & i{\neq}j,
                                   \end{cases}\\   
&\frac{\partial y_i}{\partial \phi_\gamma^j}{=}\begin{cases}
a_\gamma^i\cos{(4\pi^2\gamma f_i s(t)+\phi_\gamma^i)}, & i{=}j,\\
        0,                            & i{\neq}j,
                                   \end{cases}
\end{align*}
With $i\in\{1,2\}$, for the derivative $\nabla_\theta \hat{J}$, we also need
\begin{align*}
  &          c_i'=\sum_{\gamma=1}^{\Gamma_i} 4\pi^2 \gamma f_i a^i_\gamma\cos{(4\pi^2\gamma f_i s(t){+}\phi_\gamma^i)},\\
    &        c_i''={-}\sum_{\gamma=1}^{\Gamma_i} 16\pi^4 \gamma^2f^2_i a^i_\gamma\sin{(4\pi^2\gamma f_i s(t){+}\phi_\gamma^i)}.
\end{align*}
\begin{align*}
 &\frac{\partial c_i'}{\partial f_1}{=}\begin{cases}
\sum_{\gamma=1}^{\Gamma_i} (4\pi^2 \gamma a^i_\gamma\cos{(4\pi^2\gamma f_i s(t){+}\phi_\gamma^i)}&\\-16\pi^4 \gamma^2 a^i_\gamma \sin{(4\pi^2\gamma f_i s(t){+}\phi_\gamma^i)}), & i=1,\\
        0,                            & i=2,
                                   \end{cases}\\
&\frac{\partial c_i''}{\partial f_1}{=}\begin{cases}
\sum_{\gamma=1}^{\Gamma_i} (-32\pi^4 \gamma^2 a^i_\gamma\sin{(4\pi^2\gamma f_i s(t){+}\phi_\gamma^i)}&\\-64\pi^6 \gamma^3 a^i_\gamma s(t) \cos{(4\pi^2\gamma f_i s(t){+}\phi_\gamma^i)}), & i=1,\\
        0,                            & i=2,
                                   \end{cases}\\
                           & \frac{\partial c_i'}{\partial a^j_0}=0,\qquad \frac{\partial c_i''}{\partial a^j_0}=0,\\    
&\frac{\partial c_i'}{\partial a_\gamma^j}=\begin{cases}
\sum_{\gamma=1}^{\Gamma_i} 4\pi^2 \gamma f_i \cos{(4\pi^2\gamma f_i s(t){+}\phi_\gamma^i)}, & i=j,\\
        0,                            & i\neq j,
                                   \end{cases}\\   
&\frac{\partial c_i''}{\partial a_\gamma^j}=\begin{cases}
-\sum_{\gamma=1}^{\Gamma_i} 16\pi^4 \gamma^2 f_i^2 \sin{(4\pi^2\gamma f_i s(t){+}\phi_\gamma^i)}, & i=j,\\
        0,                            & i\neq j,
                                   \end{cases}
                                   \end{align*}
                                   \begin{align*}
&\frac{\partial c_i'}{\partial \phi_\gamma^j}=\begin{cases}
-\sum_{\gamma=1}^{\Gamma_i}4\pi^2 \gamma f_i a^i_\gamma\sin{(4\pi^2\gamma f_i s(t){+}\phi_\gamma^i)}, & i=j,\\
        0,                            & i\neq j,
                                   \end{cases}\\
&\frac{\partial c_i''}{\partial \phi_\gamma^j}=\begin{cases}
-\sum_{\gamma=1}^{\Gamma_i}16\pi^4 \gamma^2 f_i^2 a^i_\gamma\cos{(4\pi^2\gamma f_i s(t){+}\phi_\gamma^i)}, & i=j,\\
        0,                            & i\neq j.
                                   \end{cases}
\end{align*}
When the robot moves along a curve with parameter vector $\theta_-$ and detects an object at $s_-=s(t)$, for the initial parameter vector of the high-level OCP, for $i\in\{1,2\}$, we set
\begin{align*}
\forall \gamma, \phi_{\gamma}^i=4\pi^2 \gamma f_i s_-+\phi_{\gamma-}^i,\\
 a_0^i=c_i(s_-,\theta_-) -\sum_{\gamma=1}^{\Gamma_i} a_{\gamma-}^i\sin{(\phi_{\gamma}^i)},
\end{align*}
such that only a small number of constraints of the high-level OCP is violated initially and Alg.\,1 can converge fast.

\section*{Appendix C: Proofs}

\begin{proof}{(Theorem 1)}
 Since $\lambda_2$ is the only costate that depends on $u_1$ in \eqref{eq:hamil}, \eqref{eq:hamilt} implies that $\lambda^*_2(t)u^{*}_{c,p,1}(t)\leq \lambda^*_2(t)u_{c,p,1}^{}$ must hold. For $\lambda_2(t)\neq 0, t\in[0,t_2)$, we obtain $u_{p,1}(t)=-\text{sgn}(\lambda_2(t))$, i.e. $u^*_1(t)\in\{\pm 1\}$. As the input is bounded, $\lambda_2(t)=0$ can hold at isolated times, without violating \eqref{eq:hamilt}. A singular arc may exist since $\lambda_2(t)=0$ can also hold over an interval. The necessary condition for this case is obtained from the generalized first-order Legendre-Clebsch condition \cite{Choset2005}
\begin{align*}
 \frac{d^2}{dt^2} \frac{\partial \tilde{H}}{\partial u_{c,p,1}}= \ddot{\lambda}_2(t)=2c_g\alpha m_\emptyset\dot{\tilde{\bar{x}}}_2+\dot{\tilde{\bar{x}}}_3+\dot{\lambda}_1=0,\\
 -\frac{\partial}{\partial u_{c,p,1}}\left[\frac{d^2}{dt^2} \frac{\partial \tilde{H}}{\partial u_{c,p,1}}\right]\neq 0,
\end{align*}
 yielding $-1/(2c_g)$ as an additional possible input value.
\end{proof}

\begin{proof}{(Theorem 2)}
 From \eqref{eq:hamil} and $\tilde{x}_1\geq 0$, $\lambda_2(0)<0$ holds and the optimal control starts with a fragment $u_{c,p,1}^*(t)=1$ for $t\in[0,\bar{t}_1)$. Assuming that $\lambda_2(0)=a\in\mathbb{R}_{< 0}$ for the interval $[0,\bar{t}_1)$, the adjoint variable is $\lambda_2(t)=a-\frac{2c+1}{2}t^{(2)}-\lambda_1 t$ by integration of \eqref{eq:adj1}. If $\lambda_2(\bar{t}_1)=\dot{\lambda}_2(\bar{t}_1)=0$, $\lambda_1=-\sqrt{2(1+2c)a}$, the set of possible control sequences is $\{(1), (1,-1),(1,-1/2c_g),(1,-1/2c_g,-1)\}$, or in generalized form
 \begin{align}\label{eq:input1}
  u_{c,p,1}(t)=\begin{cases}
        1, &t\in[0, \bar{t}_1),\\
        -\frac{1}{2c_g}, &t\in[\bar{t}_1,\bar{t}_2),\\
        -1, &t\in[\bar{t}_2,t_2).
       \end{cases}
 \end{align}
Integrating \eqref{eq:system} with $m_q=m_{\emptyset}$ and \eqref{eq:input1}, using the switching and final conditions $\tilde{x}_1(t_2)=1,\tilde{x}_2(t_2)=0, \tilde{x}_2(\bar{t}_2)=v_{c}$ and analyzing the expressions for $\bar{t}_2=\bar{t}_1$ and $\bar{t}_2=t_2$, we obtain 
\begin{align*}
 \sqrt{\frac{2(m_{\emptyset}\alpha+c_g\alpha^2 {m_{\emptyset}}^{2}{v_c}^{2})}{2c_g+1}} \leq \bar{t}_1 \leq \sqrt{m_{\emptyset}\alpha+\frac{m_{\emptyset}^2\alpha^2v_c^2}{2}},
\end{align*}
for $0\leq v_c\leq \sqrt{\frac{2}{m_{\emptyset}\alpha}}$. Substituting the solutions of the ODE, the cost $E\{\ell\}$ becomes a function of $\bar{t}_1$ and $v_c$. Setting $\frac{\partial E\{\ell\}}{\partial \bar{t}_1}=0$, $\bar{t}^*_1\in\left\{\sqrt{\frac{2\alpha m_{\emptyset}}{2c_g+1}},\sqrt{m_{\emptyset}\alpha+\frac{\alpha^2 m_{\emptyset}^2v_c^2}{2}}\right\}$ holds. Analyzing $\frac{\partial^2 E\{\ell\}}{\partial \bar{t}_1^2}$ in the feasible interval, we can verify that the first value corresponds to the minimum. Substituting $\bar{t}^*_1$ in $E\{\ell\}$ and evaluating $\frac{\partial E\{\ell\}}{\partial v_c}$, for the optimal velocity $v^*_c\in\left\{0,\sqrt{\frac{2}{m_{\emptyset}\alpha}}\right\}$ holds. Analyzing $\frac{\partial^2 E\{\ell\}}{\partial v_c^2}$ yields $v_c^*=0$ and  $\bar{t}_2=2c_g\sqrt{\frac{2m_{\emptyset}\alpha}{2c_g+1}}$, leading to $\bar{t}_2=t_2$. Solving the ODE with \eqref{eq:input1} yields the switching point $1/(1+2 c_g)$ and the corresponding controller.
\end{proof}

\section*{Appendix D: Optimal control for two objects}

First, consider the planned path shown in Fig.\,\ref{fig:dis} with corresponding generalized trajectory shown in Fig.\,\ref{fig:dis2}. Since $o_2$ is discovered at latest when the robot reaches $s_2$, $\bar{x}_2(t_4)=0$. Thus, optimizing the trajectory for $t\in[t_4,t_6]$ can be decoupled, and the optimal control is given by
\begin{align*}
 u_c^{3*}(\bar{x})=\begin{cases}
                 1,&\text{if }\bar{x}_1\in[s_2,\frac{1+s_2}{2}),\\
                 -1,&\text{if }\bar{x}_1\in[\frac{1+s_2}{2},1).
                \end{cases}
\end{align*}
To obtain the optimal switching times of the remaining trajectory, we solve the ODE \eqref{eq:reduceddyn} with
\begin{align*}
 u_{c,p}^1(t)=\begin{cases}
                 1,&\text{if }t\in[0,t_1),\\
                 -1,&\text{if }t\in[t_1,t_2),\\
                 1,&\text{if }t\in[t_2,t_3),\\
                 -\frac{1}{2c_g},&\text{if }t\in[t_3,t_4).
                 \end{cases}
\end{align*}
With the boundary conditions $\bar{x}_1(0)=\bar{x}_2(0)=0$, and $\bar{x}_1(t_2)=s_1$, we obtain
\begin{align*}
 -\frac{(t_2-t_1)^2}{2m}+\frac{t_1}{m}(t_2-t_1)+\frac{t_1^2}{2m}=s_1,
\end{align*}
where $m=m_q\alpha$, and since $t_2\geq t_1$,
\begin{align*}
 t_1=\begin{cases}
      t_2,&\text{if } t_2\leq \sqrt{2ms_1},\\
      t_2-\frac{\sqrt{2}}{2}\sqrt{t_2^2-2ms_1}, &\text{else}.
     \end{cases}
\end{align*}
Using the boundary conditions $\bar{x}_1(t_4)=s_2, \bar{x}_2(t_4)=0$, we obtain
\small{\begin{align*}
{-}\frac{(t_4{-}t_3)^2}{4c_g m}{+}\frac{t_3{+}2t_1{-}2t_2}{m}(t_4{-}t_3){+}\frac{t_3{-}3t_2{+}4t_1}{2m}(t_3{-}t_2){+}s_1{=}s_2,\\
 t_4=(2c_g+1)t_3-4c_g(t_2-t_1),
\end{align*}}
\normalsize
yielding
\begin{align*}
 t_3{=}2(t_2-t_1)+\sqrt{\frac{{t_2}^{2}-4t_1 t_2+4{t_1}^{2}+2m(s_2-s_1)}{2c_g+1}}.
\end{align*}
Since $t_3\in\mathbb{R}_+$ and $t_3\geq t_2$, together with the derived expression for $t_1$, we obtain
\begin{align*}
 t_3=\begin{cases}
      \sqrt{\frac{t_2^2+2m(s_2-s_1)}{2c_g+1}},&\text{if } t_1=t_2,\\
      t_2,&\text{else}.
     \end{cases}
\end{align*}
From the continuity of the variables, there exists a time when $t_1=t_2=t_3=\sqrt{\frac{m}{c_g}(s_2-s_{sw})}$. Thus, the switching takes place at 
\begin{align*}
s'=\frac{1}{2c_g+1}s_2.
\end{align*}
Note that this corresponds to the result of Theorem~2, scaled for the interval $[0,s_2]$. For $t_2>\sqrt{2ms_1}$, substituting $t_1$ into the expression for $t_3=t_2$, we obtain $t_2=t_3=\sqrt{\frac{m}{c_g}(s_2-s_1)}$ and
\begin{align*}
 t_{1,s}=\sqrt{\frac{m(s_2-s_1)}{c_g}}-\frac{\sqrt{2}}{2}\sqrt{\frac{m(s_2-s_1)-2c_gms_1}{c_g}}
\end{align*}
yielding a switching at
\begin{align*}
 s''=-\frac{2^{\frac{3}{2}}\sqrt{s_2{-}s_1}\sqrt{s_2{-}(2c_g+1)s_1}{-}3s_2{+}(2c_g{+}3)s_1}{4c_g}
\end{align*}
Thus, the optimal control is given by
\begin{align}\label{optcontr1}
 u_{c,p}^{1*}(\bar{x})=\begin{cases}
                 1,&\text{if }\bar{x}_1\in[0,s_{sw,1}),\\
                 -1,&\text{if }\bar{x}_1\in[s_{sw,1},s_{sw,2}),\\
                 -\frac{1}{2c_g},&\text{if }\bar{x}_1\in[s_{sw,2},s_2),
                 \end{cases}
\end{align}
where $s_{sw,1}=\max\{s',s_1\}$ and
\begin{align*}
 s_{sw,2}=\begin{cases}
      s'', &\text{if } s_1>s',\\
      s', &\text{else.}
     \end{cases}
\end{align*}

Now consider the planned path shown in Fig.\,\ref{fig:dis1} with corresponding generalized trajectory shown in Fig.\,\ref{fig:dis3}. Analogously to the above analysis, it can be easily shown that the optimal control is 
\begin{align*}
 u_{c,p}^{1*}(\bar{x})=\begin{cases}
                 1,&\text{if }\bar{x}_1\in[0,s_{sw,1}),\\
                 -1,&\text{if }\bar{x}_1\in[s_{sw,1},s_{sw,2}),\\
                 -\frac{1}{2c_g},&\text{if }\bar{x}_1\in[s_{sw,2},s_{sw,3}),\\
                  1,&\text{if }\bar{x}_1\in[s_{sw,3},s_{sw,4}),\\
                 -1,&\text{if }\bar{x}_1\in[s_{sw,4},s_{sw,5}),\\
                 -\frac{1}{2c_g},&\text{if }\bar{x}_1\in[s_{sw,5},s_2),
                 \end{cases}
\end{align*}
where with $\bar{x}_2(t_4)=\bar{x}_2(t_6)$ we obtain the appropriate switching spots $s_{sw,1}$ to $s_{sw,5}$. Thus, for $n$ unexplored intervals in $[0,1]$, it can be shown that the optimal control consists of a string of $n$ controls of the form \eqref{optcontr1} with appropriate switching conditions.

\bibliographystyle{IEEEtran}
\bibliography{literature}
\end{document}